\documentclass[twocolumn, twocolappendix]{aastex631}
\usepackage{amsmath}
\usepackage{bm}

\def\Mpc{{\rm\thinspace Mpc}}

\begin{document}

%\title{Zooming by in the CARPool(GP) lane: Extending CAMELS-TNG simulations to the group-mass scale across a 29-dimensional parameter space}
\title{Zooming by in the CARPoolGP lane: new CAMELS-TNG simulations of zoomed-in massive halos}

\author[0000-0002-2318-3087]{Max E. Lee}
\affiliation{Department of Astronomy, Columbia University, 538 West 120th Street, New York, NY 10027, USA}

\author[0000-0002-3185-1540]{Shy Genel}
\affiliation{Center for Computational Astrophysics, Flatiron Institute, 162 Fifth Ave, New York, NY, 10010, USA}
\affiliation{Columbia Astrophysics Laboratory, Columbia University, 550 West 120th Street, New York, NY 10027, USA}

\author[0000-0002-5854-8269]{Benjamin D. Wandelt}
\affiliation{CNRS \& Sorbonne Universit\'{e}, Institut d’Astrophysique de Paris (IAP), UMR 7095, 98 bis bd Arago, F-75014 Paris, France}
\affiliation{Center for Computational Astrophysics, Flatiron Institute, 162 Fifth Ave, New York, NY, 10010, USA}

\author{Benjamin Zhang}
\affiliation{Ward Melville High School, 380 Old Town Road, East Setauket, NY 11733, USA}

\author{Ana Maria Delgado}
\affiliation{Center for Astrophysics $\vert$ Harvard \& Smithsonian, 60 Garden Street, Cambridge, MA 02138, USA}

\author{Shivam Pandey}
\affiliation{Columbia Astrophysics Laboratory, Columbia University, 550 West 120th Street, New York, NY 10027, USA}

\author[0000-0001-8914-8885]{Erwin T.\ Lau}
\affiliation{Center for Astrophysics $\vert$ Harvard \& Smithsonian, 60 Garden Street, Cambridge, MA 02138, USA}

\author[0000-0002-5840-0424]{Christopher Carr}
\affiliation{Columbia University Department of Astronomy, 538 West 120th Street, New York, NY 10027, USA}

\author[0000-0001-7163-8712]{Harrison Cook}
\affiliation{New Mexico State University, Department of Astronomy, PO Box 30001 MSC 4500, Las Cruces, NM 88001, USA}

\author[0000-0002-6766-5942]{Daisuke Nagai}
\affiliation{Department of Physics, Yale University, New Haven, CT 06520, USA}

\author[0000-0001-5769-4945]{Daniel Angles-Alcazar}
\affiliation{Department of Physics, University of Connecticut, 196 Auditorium Road, U-3046, Storrs, CT, 06269, USA}
\affiliation{Center for Computational Astrophysics, Flatiron Institute, 162 Fifth Ave, New York, NY, 10010, USA}

 \author{Francisco Villaescusa-Navarro}
\affiliation{Center for Computational Astrophysics, Flatiron Institute, 162 Fifth Ave, New York, NY, 10010, USA}

\author{Greg L. Bryan}
\affiliation{Columbia University Department of Astronomy, 538 West 120th Street, New York, NY 10027, USA}
\affiliation{Center for Computational Astrophysics, Flatiron Institute, 162 Fifth Ave, New York, NY, 10010, USA}

...



\begin{abstract}
Galaxy formation models within cosmological hydrodynamical simulations contain numerous parameters with non-trivial influences over the resulting properties of simulated cosmic structures and galaxy populations. It is computationally challenging to sample these high dimensional parameter spaces with simulations, particularly for halos in the high-mass end of the mass function. In this work, we develop a novel sampling and reduced variance regression method, \textit{CARPoolGP}, which leverages built-in correlations between samples in different locations of high dimensional parameter spaces to provide an efficient way to explore parameter space and generate low variance emulations of summary statistics. We use this method to extend the Cosmology and Astrophysics with MachinE Learning Simulations (CAMELS) to include a set of 768 zoom-in simulations of halos in the mass range of $10^{13} - 10^{14.5} M_\odot\,h^{-1}$ that span a 28-dimensional parameter space in the IllustrisTNG model. With these simulations and the CARPoolGP emulation method, we explore parameter trends in the Compton $Y-M$, black hole mass-halo mass, and metallicity-mass relations, as well as thermodynamic profiles and quenched fractions of satellite galaxies. We use these emulations to provide a physical picture of the complex interplay between supernova and active galactic nuclei feedback. We then use emulations of the $Y-M$ relation of massive halos to perform Fisher forecasts on astrophysical parameters for future Sunyaev-Zeldovich observations and find a significant improvement in forecasted constraints. We publicly release both the simulation suite and CARPoolGP software package.
\end{abstract}


%% automatically renumber them.https://www.overleaf.com/project/64ad8a614d24f91315142e43

\section{Introduction} \label{sec:intro}

A wide range of cosmological surveys are coming soon, including cosmic microwave background (CMB) experiments such as the Simons Observatory \citep{SimonsObservatory} and CMB-S4 \citep{cmbs4}, galaxy surveys such as the Legacy Survey of Space and Time (LSST) at the Vera Rubin Observatory \citep{LSST} with EUCLID \citep{Euclid}, and the Nancy Grace Roman space observatory \citep{Roman}, X-ray observatories such as e-Rosita \citep{Erosita}, and radio interferometers such as SKA \citep{SKA}. These new surveys have the potential to provide solutions to long-standing questions in cosmology. However, uncertainties in baryonic processes, such as active galactic nuclei and supernova feedback, will limit the theoretical interpretations of these future observations. The small-scale systematics these processes introduce, which have previously been a subdominant effect in cosmological analyses, will now be a prevalent source of uncertainty. So, it has become necessary to develop methods for modeling them.

The most straightforward treatment of baryonic processes is to remove small-scale data affected by baryonic systematics directly. This method has indeed been used, for example, in weak lensing analysis with the Dark Energy Survey (DES) \citep{Secco-2022}, the latest KiloDegree Survey (KiDs) \citep{Li-2023} and more recently in HyperSuprimeCam (HSC) \citep{Li-2023_HSC}. However, scale cuts must be sufficiently large to remove baryonic processes, resulting in loss of information.

Instead of a scale cut, one could model baryonic effects using cosmological hydrodynamical simulations \citep{Vogelsberger-2014,  Crain-2015, Pillepich-2018, Dave-2019}. Scale cuts offers an advantage that the small, non-linear scales, shown to contain a wealth of cosmological parameter constraining power \citep{Seljak-2017, FNV2021a, FVN2021b, Natali2023}, can be incorporated into the analysis. However, even these simulations must approximate physical effects on scales below the resolution of the simulation grid (subgrid physics), which is uncertain and handled differently across simulations \citep{VanDaalen-2014,  Vogelsberger-2020, Villaescusa-Navarro-2021}. These differences reflect current uncertainties in feedback prescriptions and galaxy formation in general. 

The high computational cost associated with hydrodynamical simulations limits their use in approaches that attempt inference through forward modeling. However, leveraging advances in Machine Learning (ML) can provide an avenue for marginalizing uncertainty in baryonic effects by developing neural networks \citep{Ribli-2019, Villaescusa-2022, Lu-2023}. An essential requirement of such a task is some data set to train machine learning models. The Cosmology and Astrophysics with MachinE Learning Simulations (CAMELS)\footnote{\url{https://www.camel-simulations.org/}} \citep{Villaescusa-Navarro-2021}, are pursuing the generation of such a training set with state-of-the-art cosmological hydrodynamical simulations spanning multiple subgrid prescriptions including IllustrisTNG \citep{Weinberger-2017, Pillepich-2018}, SIMBA \citep{Dave-2019}, and Astrid \citep{Ni-2022, Bird-2022}. Thousands of such simulations have already been created and have been shown to be helpful in a wide variety of astrophysical and cosmological contexts, ranging from improvements in galaxy cluster scaling relations \citep{Amodeo-2021, Wadekar-2023a, Wadekar-2023b}, to exploring the impact of active galactic nuclei (AGN) and supernova (SN) feedback on galaxy clustering \citep{Gebhardt-2023, Delgado-2023}.

All existing CAMELS simulations have in common that they encompass a volume of $(25\, \Mpc\,h^{-1})^3$. Irrespective of cosmological or astrophysical parameters, this significantly limits the halo mass range in these volumes, with only around half of the simulations containing any halos with $ M > 10^{13.5}\, M_\odot\,h^{-1}$ and $15\%$ containing any halos with $M > 10^{14.5}\, M_\odot\,h^{-1}$. The range of masses above $10^{13}\,M_\odot \, h^{-1}$ is particularly interesting for future cosmology and astrophysics studies as it will be observed through secondary CMB anisotropies such as the thermal and kinetic Sunyaev-Zeldovic effects. These observations can provide stringent constraints on the matter density of the universe \citep{2002ARA&A..40..643C}, the amplitude of matter fluctuations \citep{Komatsu-2001, Seljak-2017}, the sum of the neutrino masses \citep{Madhavacheril-2017}, and the dark energy equation of state. Furthermore, observations of the SZ effect can constrain astrophysical processes occurring in galaxy groups and clusters and their thermodynamic properties \citep[][for review]{Battaglia-2017, Schaan-2021, Moser-2022, Hadzhiyska-2023b, Mroczkowski2019}. Similarly, observed X-ray emission and spectra from galaxy groups and clusters can aid in constraining the complex feedback and formation processes in the multiphase intracluster medium (ICM) and Intragroup Medium (IGrM) \citep{Singh-2018, Tyler-2021, Singh-2022, Lee-2022, lovisari-2023}.

In fact, attempts to constrain both galaxy formation and cosmology using multi-wavelength observations, improved simulations, and analysis algorithms are currently ongoing, for example, in the Learning the Universe collaboration\footnote{\url{https://www.learning-the-universe.org/}}. However, for future analyses to proceed, increasing the mass range of simulated models to include group- and cluster-scale halos is imperative.

In this work, we seek to do just this. Our main goal is to generate a set of galaxy group to cluster scale zoom-in simulations that span a wide range of astrophysical and cosmological parameters in the IllustrisTNG model's parameter space. We design this suite so that it can be used to train an emulator to predict halo quantities (such as the integrated Compton $Y_{SZ}$ signal of the tSZ effect) throughout the entire parameter space. In this way, the emulated halo quantities can easily extend previous CAMELS experiments at the high-mass end of the mass function, opening up the possibility of future analyses that are focused independently on the effects of more massive halos\footnote{We publicly release all simulations. For access see \url{zoomgz.readthedocs.io}}. 

This task is non-trivial due to the roles played by sample variance and high parameter space dimensionality. Unlike the $(25\, \Mpc\,h^{-1})^3$ CAMELS boxes, where there are typically multiple halos at each mass and parameter space location (below the group mass scale), in this work, we will be limited to significantly fewer halos due to the high computational cost of simulating massive objects, leading to results that are more vulnerable to sampling effects. Similarly, generating a simulation suite to cover such a high-dimensional parameter space introduces the curse of dimensionality. Often, a Latin Hypercube or Sobol Sequence \citep{Sobol-1967} is used to sample high-dimensional spaces to best capture variations throughout the entire parameter space. However, whether either of these sampling approaches is optimal for generating a training set for an emulator remains to be determined.

Irrespective of the sampling strategy, it is imperative that we minimize the uncertainty introduced by the small number of groups and clusters we can simulate. In this work, we introduce the novel reduced variance emulator CARPoolGP with an associated active learning parameter sampling method, which builds on the success of the CARPool reduced variance estimator \citep{Chartier-2021,Chartier-2022a,Chartier-2022b} and generalizes the approach to a broad parameter space. We then apply our novel method to the 28-dimensional IllustrisTNG galaxy formation model to build a new suite of 768 galaxy groups and clusters.

The organization of this work is as follows. In Section~\ref{sec:methods}, we introduce the most general formalism for the CARPoolGP emulation and sampling approach and provide a one-dimensional example to demonstrate its efficacy. We then discuss the setup of the cosmological zoom-in simulation suite in Section~\ref{sec:simulations} and how we have applied the CARPoolGP methodology to this case. We present simple applications of the zoom-in suite and CARPoolGP by emulating various scaling relations, including the Compton $Y-M$ relation in Section~\ref{sec:results}. We then perform a Fisher forecast on astrophysical parameters using the derivatives of the emulated $Y-M$ relation and discuss caveats to our testing methods in Section~\ref{sec:discussion}. We conclude with the key takeaways from this work in Section~\ref{sec:conclusion}.

\section{Sampling and emulating with CARPoolGP}\label{sec:methods}
Many situations arise where one is interested in some functional form of the parameter space dependence of a quantity but is limited to a few data points riddled with sample variance. For example, it would naively require a diverse population of halos across numerous masses to predict how the circumgalactic medium temperature changes as a function of halo mass.
However, the generation of samples may be expensive, which in turn could stymie attempts at obtaining a low-variance prediction of the quantity's parameter dependence. It is then advantageous to develop an emulator that provides not only the quantity of interest but also a measurement of the associated uncertainty. One general solution to this problem is using Gaussian process regression, which treats each sample at a location in parameter space as a random process drawn from a distribution that can be learned through regression. In this way, one can extract the correlations between quantities throughout the parameter space and the mean parameter dependence on the quantities (see \citet{Bird-2019, Aigrain-2022, Bird-2022, Ho-2022} for a review of the Gaussian process regression and applications).

Correlations in parameter space can often be constructed between samples, as in the case of cosmological simulations, where the initial conditions can be controlled. This work seeks to improve upon general Gaussian process regression in situations such as these, where correlations naturally exist or can be generated between samples. In this first section, we introduce a novel sampling approach and an associated emulator called CARPoolGP\footnote{\url{github.com/Maxelee/CARPoolGP}} \citep{Lee-2024} to do exactly this. 

We begin the section by developing a general formalism for the method. We then describe an improvement to the sampling scheme, and finally, we end the section with a simple, one-dimensional toy example that demonstrates the power of CARPoolGP. We provide a consolidated summary of variables, their definitions, and the equations where they are introduced in Table~\ref{tab:my_label} for easy reference.

\subsection{Gaussian Process general formalism}\label{sec:CPTheory}

\begin{table}
\centering
\begin{tabular}{|p{0.15\linewidth}|p{0.6\linewidth}|p{0.1\linewidth}|}
\hline
 Variable     &   Meaning & Eq. \\
 \hline
 $\bm{\theta}_{i}$ & Parameter vector for sample $i$ & Eq.~\ref{eq:tilde_Q}\\
$\tilde{Q}$   &  Noisy quantity extracted at a parameter space location & Eq.~\ref{eq:tilde_Q}\\
Q & True value of the quantity at a parameter space location & Eq.~\ref{eq:tilde_Q}\\
$\epsilon$ & noise associated with a quantity & Eq.~\ref{eq:tilde_Q}\\
$\sigma^2_Q$ & Level of noise to determine $\epsilon$ & Eq.~\ref{eq:tilde_Q}\\
$C_{ij}$ & Covariance matrix for base quantities & Eq.~\ref{eq:cov_C}\\
$V_{ij}$ & Smooth noise kernel for base quantities & Eq.~\ref{eq:cov_C}\\
$F_{ij}$ & Sample variance component for base quantities & Eq.~\ref{eq:cov_C}\\
$\mu_Q$ & Prior on base quantities & Eq.~\ref{eq:liklihood}\\
$\tau$ & Vector of hyperparameters for noise kernels & Eq.~\ref{eq:liklihood}\\
$\bm{\theta}_{i}^\ast$ & Parameter vector to be emulated & Eq.~\ref{eq:prediction}\\
$K_\ast$ & Predictive covariance between emulated and existing points & Eq.~\ref{eq:prediction}\\
$K_{\ast\ast}$ & Predictive covariance between emulated points & Eq.~\ref{eq:prediction}\\
$\tilde{Q}^S$ & Noisy surrogate quantity  & Eq.~\ref{eq:cov_D}\\
$D_{ij}$ & Covariance matrix for surrogate quantities & Eq.~\ref{eq:cov_D}\\
$W_{ij}$ & Smooth noise kernel for surrogate quantities& Eq.~\ref{eq:cov_D}\\
$E_{ij}$ & Sample variance noise component for surrogate quantities& Eq.~\ref{eq:cov_D}\\
$X_{i j}$ & Covariance between base and surrogate Quantities & Eq.~\ref{eq:cov_X}\\
$Y_{i j}$ &  Smooth noise kernel for base and surrogate covariance& Eq.~\ref{eq:cov_X}\\
$M_{i j}$ &  Covariance kernel for sample variance between base and surrogates& Eq.~\ref{eq:cov_X}\\
$\mu_R$ & Prior on surrogate quantities & Eq.~\ref{eq:CPliklihood}\\
\hline
\end{tabular}
\caption{Summary of variables used in developing CARPoolGP from Section~\ref{sec:CPTheory} and Section~\ref{sec:CARPool} into a table for reference. }
\label{tab:my_label}
\end{table}
Consider some quantity $Q$  we want to fit as a function of parameter space locations $\bm{\theta}_i$ using noisy samples $\tilde{Q}$ with sampling error $\epsilon$. The quantity $Q$ and samples $\tilde{Q}$ are then related through
\begin{equation}\label{eq:tilde_Q}
    \tilde{Q}(\bm{\theta}_{i}) = Q(\bm{\theta}_{i}) + \epsilon(\bm{\theta}_{i}),
\end{equation}
where we take $\epsilon$ to have $\langle\epsilon\rangle=0$ and $\langle\epsilon^2\rangle=\sigma_Q^2(\bm{\theta}_{i})$ and $\sigma_Q^2(\bm{\theta}_{i})$ is the variance level on the noisy quantity that can depend on location in parameter space.  

On the one hand, one typically wants to average over multiple $\tilde{Q}(\bm{\theta}_{i})$, each with a different noise realization, to obtain a good estimate of $Q(\bm{\theta}_{i})$. Using a fitting procedure, new parameter points can be emulated with some level of predictive variance on the noiseless quantity $Q(\bm{\theta}_{i})$. Unfortunately, if the cost of producing a single noisy measurement $\tilde{Q}(\bm{\theta}_{i})$ is high, as is often the case, this procedure is difficult and can lead to high variance and poor accuracy of the emulated quantity. Our task is then to answer: \textit{How do we achieve an emulation with a low predictive variance of a quantity across a parameter space, when we can only obtain a small number of noisy measurements of the quantity at each of a limited set of parameter locations?}

With Gaussian process regression, we imagine that $N_B$ samples of some noisy quantity $\tilde{Q}(\bm{\theta}_{i})$ exist throughout the parameter space. We denote this set $B\equiv\{\bm{\theta}_{i}| i=1, 2, ..., N_B\}$ by the \textit{base} samples (for reasons that will soon become clear). Base samples have a covariance matrix,
\begin{equation}\label{eq:cov_C}
    C_{ij} = \text{Covar} \left(\tilde{Q}(\bm{\theta}_{i}), \tilde{Q}(\bm{\theta}_{j})\right) = V(\bm{\theta}_{i} \,, \bm{\theta}_{j}) + F(\bm{\theta}_{i}, \bm{\theta}_{j}),
\end{equation}
that contains two matrix terms, $V$, which represents the smooth and systematic variation in the quantity $Q(\bm{\theta}_{i})$ as a function of parameters, and $F$, which contains the sample fluctuations stemming from the noise. In this way, $V$ can be considered to be the ``true" covariance associated with $Q$, while $F$ is some error associated with the covariance, which stems from the fact that our measurements are not of $Q$ but rather of $\tilde{Q}$. 

Both matrices encode correlations between individual samples in their systematic and sample variances, respectively, but in the typical case of independent samples with uncorrelated sample variance, $F$ can be written in a less general form such that $F_{ij} = \sigma_{Q, i}^2\mathcal{I}$ where $\mathcal{I}$ is the identity matrix. We have absorbed the parameter dependence into the subscripts of our matrices for brevity such that the parameter indices become matrix indices (e.g., $V(\bm{\theta}_{i} \,, \bm{\theta}_{j}) \rightarrow V_{ij}$). The noise matrices are any arbitrary linear combination of Gaussian kernel functions, such as Radial Basis functions (RBF), Matern, or linear exponential kernels (see Ch. 4 of \citet{Ramussen-2006} for more kernels). Kernel functions contain free hyperparameters, $\bm{\tau}$, and depend on the distance between the parameter points. As an example, an RBF would have the form of: 
\begin{equation}\label{eq:rbf}
    V_{\rm RBF}(\alpha, \bm{\gamma}, \bm{\theta}_{i}, \bm{\theta}_{j}) = \alpha \prod_{p} \exp\left(-\gamma_p \, d_E(\bm{\theta}_{i}, \bm{\theta}_{j})^2\right),
\end{equation}
where $d_E(\cdot)$ is the Euclidean distance between the parameter points, $\bm{\gamma}$ is a vector of smoothing scale factors, and $\alpha$ is an amplitude constant. The product comprises $N_p$ kernels where $N_p$ is the dimensionality of the parameter space, and $p$ is the index over a given parameter, each containing unique smoothing scale factors. We can consider arbitrarily complex kernel functions with increasing numbers in the hyperparameter vector $\bm{\tau}$.

Given prior knowledge of a quantity's variation, we can impose the prior, $\mu_Q(\bm{\theta}_{i})$ on the mean function, and use the Gaussian likelihood function,
\begin{equation}\label{eq:liklihood}
\begin{split}
    \mathcal{L}(\tau) &= \frac{1}{(2\pi)^{N_B/2}} |C_{ij}(\bm{\tau})|^{-1/2}\,\times \\
    &\exp\left(-\frac{1}{2}(\tilde{\bm{Q}}-\bm{\mu}_Q)^T C_{ij}(\bm{\tau})^{-1}(\tilde{\bm{Q}} - \bm{\mu}_Q)\right),
\end{split}
\end{equation}
which depends on the vector of hyperparameters, $\bm{\tau}$, used to generate the kernel matrices, and $N_B$, the number of parameters in the set $B$. An optimal set of hyperparameters, $\hat{\bm{\tau}}$ can be found through an optimization method that maximizes the Eq.~\ref{eq:liklihood} such as Stochastic Gradient Descent (SGD). We impose prior bounds on the parameters in $\bm{\tau}$ during optimization such that the hyperparameters controlling scale and amplitude for a given Gaussian kernel (e.g. $\gamma_p$ and $\alpha$) can span between [0,$\infty$), and the mean function, $\mu_Q$ can span $(-\infty, \infty)$. $\hat{\bm{\tau}}$ can then be used to generate an estimator for $Q(\bm{\theta}_{i}^\ast)$ and its predictive variance $\sigma^2(\bm{\theta}_{i}^\ast)$, at any new set of parameter values, $B^\ast\equiv\{\bm{\theta}_{i}^\ast|; i=1,2,\dots,M\}$, $B^\ast \notin B$, as, 
\begin{equation}\label{eq:prediction}
\begin{split}
    Q(\bm{\theta}_{i}^\ast)  & = \text{K}_\ast\, C^{-1} \left(\tilde{Q}(\bm{\theta}_{i})-\mu_Q\right) + \mu_Q\\
    \sigma^2(\bm{\theta}_{i}^\ast) & = \text{K}_{\ast\ast} - \text{K}_\ast C^{-1}\text{K}_\ast^T,
\end{split}
\end{equation}
with $K_\ast = V(\bm{\theta}_{i}^\ast, \bm{\theta}_{j} ; \hat{\bm{\tau}})$, and $K_{\ast\ast} = V(\bm{\theta}_{i}^\ast, \bm{\theta}_{i}^\ast ; \hat{\bm{\tau}})$, both evaluated using the set of optimized hyperparameters, $\hat{\bm{\tau}}$.

To be clear, the quantity $\sigma^2({\bm{\theta}_i^\ast})$ is the variance associated with the prediction by the emulator of the mean $Q(\bm{\theta}_{i}^\ast)$ at one parameter location. But when predicting the mean at multiple parameter points, we obtain the predictive covariance: $\sigma^2({\bm{\theta}_{i}^\ast, \bm{\theta}_{j}^\ast})$. Throughout the text, when making predictions on multiple values, we will refer to the diagonal of the predictive covariance matrix as the predictive variance and the square root of the diagonal $\sigma({\bm{\theta}_{i}^\ast,\bm{\theta}_i^\ast})$ as the predictive uncertainty. These quantities are distinct from the \textit{sample} variance, $\sigma_Q^2$, representing intrinsic scatter associated with a quantity, which can, in general, be a hyperparameter learned through regression.

So far, we have described a brief and general formalism for typical Gaussian process regression. For a more detailed description, we recommend \cite{Aigrain-2022} or \citet{Ramussen-2006}.

\subsection{Introducing CARPoolGP} \label{sec:CARPool}
\begin{figure}[t]
    \centering
    \includegraphics[width=0.48\textwidth]{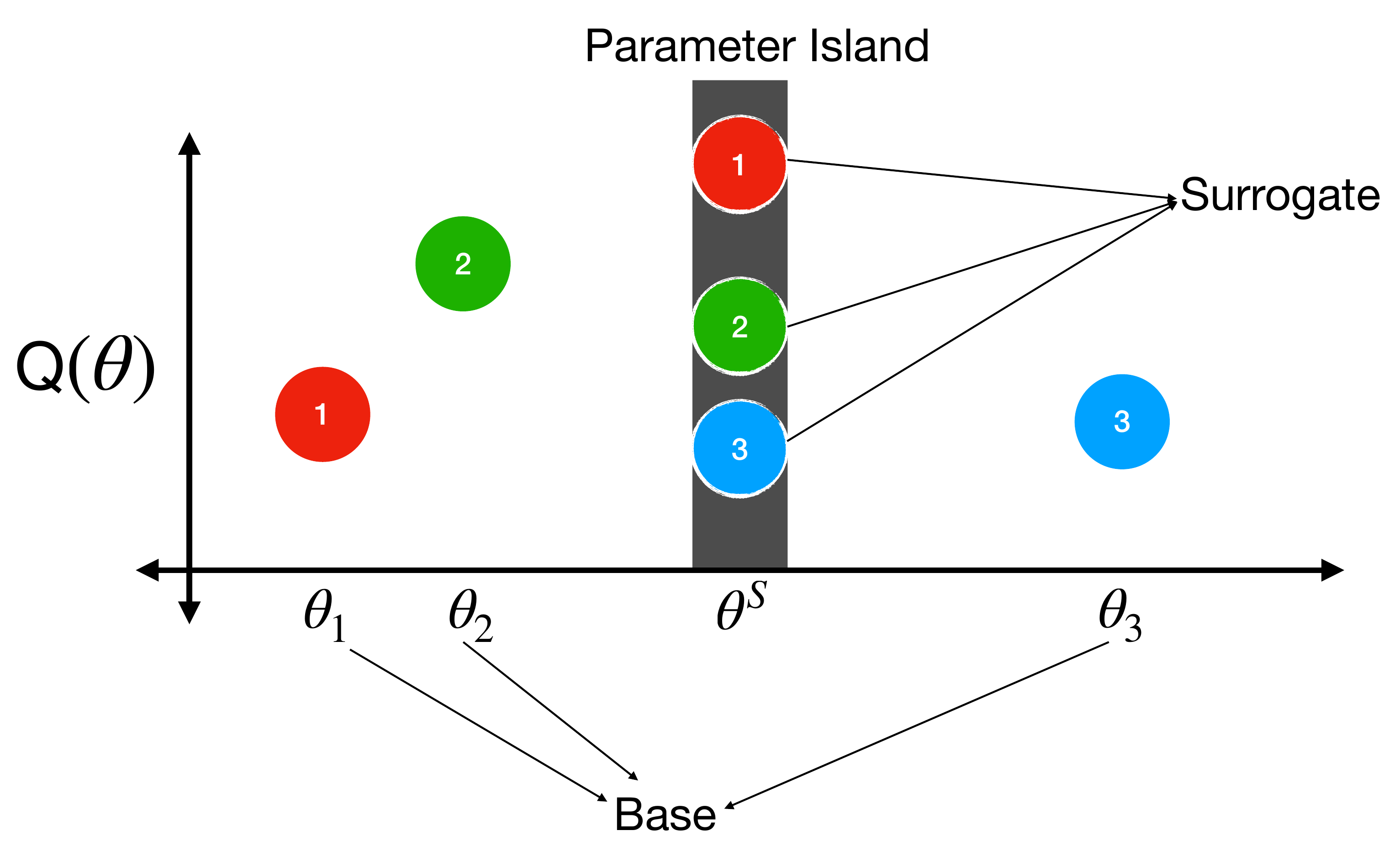}
    \caption{A cartoon including the CARPoolGP components. The base samples are spread over the parameter space at $\theta_{i}$, while the surrogates live on a parameter island at $\theta^S$. A base surrogate pair can be identified by the duo that has matching sample noise realization, denoted by the number in the circle. In this simple cartoon, the surrogates on the parameter island can broadcast their knowledge of their error to base samples represented here by the colors (e.g., red samples are above the mean at the parameter island, blue samples are below the mean at the parameter island).}
    \label{fig:cartoon}
\end{figure}

CARPoolGP builds upon this general framework of Gaussian process regression while borrowing the key principle of the CARPool method to reduce variance in a predicted quantity \citep{Chartier-2021}. CARPool achieves this through an experimental design that leverages correlations between `expensive' \textit{base} samples and `cheap' \textit{surrogate} samples. In CARPool, numerous inexpensive surrogate samples can be generated at arbitrarily low costs to accurately estimate the mean of the surrogate quantity. The correlation between surrogates and a few base samples is then leveraged to apply this accurate estimate to the base quantity. A reduced variance and accurate expression of some mean quantity can be constructed without the need for many expensive samples. CARPool has been adopted for mock catalog generation in DESI \citep{Ding-2022}, and similar work has explored sampling and emulation using correlated simulations at different fidelity such as \citet{Ho-2022}

The novelty of CARPoolGP is in applying this reduced variance procedure across an entire parameter space. In CARPoolGP, base samples provide parameter space coverage at more parameter space locations but are expected to be riddled with high variance. Surrogates, however, can be generated with the same or different process as the base (e.g. different cost), and are generated at an alternate set of locations that we call \textit{parameter islands}. Parameter islands contain multiple surrogates, providing a reduced variance estimate at these locations. By correlating isolated base samples to surrogates, the parameter space can be sampled more densely while borrowing the variance reduction achieved at parameter islands. In Fig.~\ref{fig:cartoon}, we present a simplistic picture of the CARPoolGP sampling method. We show three base samples spread through a one-dimensional parameter space with some associated noisy quantity $\tilde{Q}(\theta_i)$. Surrogate samples live on the parameter island, which is shown as the black bar, located at $\theta^S$. A base-surrogate pair, which has correlated sample noise realizations, is shown by samples with matching numbers. The colors show how surrogates from the parameter island can maintain some conception of their variation at the parameter island (e.g., red is too high, blue is too low) and broadcast this to the base samples.

Parameter islands have some set of parameter space locations $S\equiv\{\theta_{i}|i=1, 2, ..., N_S\}$, $S \notin B$, where $N_S$ is the total number of parameter islands. Surrogates are unique in that they can have their sample variance correlated with a base sample, and typically, there will be numerous surrogates on an individual island. In CARPoolGP, surrogate samples can be generated with varying cost-based processes, as done in \cite{Chartier-2021} or with the same process as base samples (e.g., same cost). The advantage of this flexibility will be clear when we are interested in the evolution of a quantity across a parameter space for a particular process and assume that locations in this space contain correlations. Then, with surrogate quantities generated by the same process as base quantities (e.g., using the same cost for generation), one can leverage the variance reduction at surrogate locations and broadcast this via correlations at base locations.

Similarly to base quantities, we will call the quantities we measure from the surrogates $\tilde{Q}^S$, which have sample noise $\epsilon_S$, 
\begin{equation}
    \tilde{Q}^S = Q^S + \epsilon_S.
\end{equation}
In principle, $Q^S$ and $Q$ can arise from different functions, and $\epsilon_S$ can differ from $\epsilon$. This scenario could arise in the situation where surrogate quantities are drawn from processes with different costs, e.g., expensive versus fast simulations \citep{Chartier-2021}. However, the surrogates may also be produced by the same process as the base samples, which indeed is the specific implementation in this work, as described below. Therefore, in the examples used in this work, the functions that generate $Q$ and $Q^S$ are the same. Similar to the bases, surrogates have some covariance matrix,
\begin{equation}\label{eq:cov_D}
    D_{ij} = \text{Covar} \left(\tilde{Q}^S_{i}, \tilde{Q}^S_{j}\right) = W_{ij}+ E_{ij},
\end{equation}
where $W_{ij}$ is again the smoothly varying portion of the quantity's variance in analogy to $V_{ij}$ in Eq.~\ref{eq:cov_C}. Unlike the $F_{ij}$ matrix in Eq.~\ref{eq:cov_C}, which is typically diagonal for independent samples, the $E_{ij}$ matrix does not need to be diagonal. Instead, off-diagonal elements can exist if correlations exist between the surrogate simulations at different islands. For example, if surrogates from different islands in parameter space share the same realization of the sample noise, then the off-diagonal elements $E_{i\neq j}$ would contain values of $r_{ij}\sigma_{i}\sigma_{j}$, where $r_{ij}$ represents the correlation between the two surrogates with sample variances $\sigma_i = \sigma_S(\bm{\theta}_{i})$ and $\sigma_j=\sigma_S(\bm{\theta}_{j})$.

Assuming again that the variance among surrogates is Gaussian and can be assigned some prior mean term $\mu_S$, then the surrogate and base samples can be represented as normally.
\begin{equation}
    \begin{split}
        \begin{pmatrix}
        \tilde{Q}\\
        \tilde{Q}^S
        \end{pmatrix}
        \sim \mathcal{N}
        \left(\begin{pmatrix}
            \mu_Q\\
            \mu_S
        \end{pmatrix}, 
        \begin{pmatrix}
            C_{ij} \,\,\, X_{ij}\\
            X_{ij}^T\, D_{ij}
        \end{pmatrix}\right),
    \end{split}
\end{equation}
where $X_{ij}$ is the covariance between the base and surrogate samples.  These, in turn, can again be modeled with a combination of both smooth and sample covariance kernels,  
\begin{equation}\label{eq:cov_X}
       X_{ij} = \text{Covar}\left(\tilde{Q}_{i}, \tilde{Q}^S_{j}\right) =  Y_{ij} + M_{ij},
\end{equation}
where $Y_{ij}$ and $M_{ij}$ should be considered analogs to those shown in Eq.~\ref{eq:cov_C} and Eq.~\ref{eq:cov_D}. However, care must be taken in their construction. First, the most general form of $Y_{ij}$ should allow the smooth functions $Q(\bm{\theta}_{i})$ and $Q^S(\bm{\theta}_{i})$ to be different in which case the distance between the location of the parameters in $B$ and $S$ must be modified in such a scenario to account for the differences in functional form. We account for this modulation with the learnable hyperparameter $\Delta q_{BS}$, which can be linearly incorporated into the distance calculation with 
\begin{equation}\label{eq:d_Y}
    d_Y(\bm{\theta}_{i}, \bm{\theta}_{j}) = d_E(\bm{\theta}_{i}, \bm{\theta}_{j}) + \Delta q_{BS}.
\end{equation}
The $M_{ij}$ matrix is similar to the $F_{ij}$ and $E_{ij}$ matrices in Eqs.~\ref{eq:cov_C} and~\ref{eq:cov_D}, except the $M_{ij}$ matrix considers the correlation between base and surrogates, and can be non-zero for any base-surrogate pair generated to have correlated sample variance. 

This correlation is the critical component of the CARPoolGP method, and without it, we do not expect an increased performance compared to a standard Gaussian process with an equal number of samples. With a high level of correlation between base and surrogate samples through $M_{ij}$, both base and surrogate can independently calibrate themselves on nearby parameter samples from multiple locations in the parameter space.

Writing the block covariance matrix in a compact form as
\begin{equation}\label{eq:sigma}
    \Sigma_{ij} = \begin{pmatrix}
            C_{ij} \,\,\, X_{ij}\\
            X_{ij}^T\, D_{ij}
        \end{pmatrix},
\end{equation}
allows us to rewrite the likelihood function of Eq.~\ref{eq:liklihood} for the general CARPoolGP case as,
\begin{equation}\label{eq:CPliklihood}
\begin{split}
    \mathcal{L}(\tau) = &\frac{1}{(2\pi)^{N/2}} |\Sigma(\tau)|^{-1/2}\,\times\\
    &\exp\left(-\frac{1}{2}\begin{pmatrix}
        \tilde{Q}-\mu_Q\\
        \tilde{Q}^S-\mu_S
        \end{pmatrix}^T \Sigma(\tau)^{-1}\begin{pmatrix}
        \tilde{Q}-\mu_Q\\
        \tilde{Q}^S-\mu_S
        \end{pmatrix}\right).
\end{split}
\end{equation}
It should be noted that the most general form of $\Sigma_{ij}$ allows the generation of multiple surrogates for each base, for example, in the scenario where a base sample has surrogates drawn from multiple different cost-varying processes. The effect on $\Sigma_{ij}$ would be the addition of kernel matrices analogous to $D_{ij}$ along the diagonal of the block matrix, $X_{ij}$ along the off diagonals with an associated hyperparameter $\Delta q_{BS}$, and a new set of kernel matrices that describe the correlations between surrogates of different processes to extend the off diagonals of the block matrix. 

By maximizing the likelihood function of Eq.~\ref{eq:CPliklihood} with respect to $\tau$, and using the optimized $\Sigma(\hat{\bm{\tau}})$ as the covariance matrix, one can again build an estimator as in Eq.~\ref{eq:prediction} for the quantity of interest at new points in parameter space, 
\begin{equation}\label{eq:CPprediction}
\begin{split}
     Q(\bm{\theta}_{i}^\ast)  & = \text{K}_\ast(\hat{\bm{\tau}})\, \Sigma^{-1}_{ij}(\hat{\bm{\tau}}) \begin{pmatrix}
        \tilde{Q}-\mu_Q\\
        \tilde{Q}^S-\mu_S
        \end{pmatrix} + \begin{pmatrix}
        \mu_Q\\
        \mu_S
        \end{pmatrix}\\
    \sigma^2(\bm{\theta}_{i}^\ast, \bm{\theta}_{j}^\ast) & = \text{K}_{\ast\ast}(\hat{\bm{\tau}}) - \text{K}_\ast(\hat{\bm{\tau}})\Sigma_{ij}^{-1}(\hat{\bm{\tau}})\text{K}_\ast^T(\hat{\bm{\tau}}),
\end{split}
\end{equation}
with the predictive kernels defined as,
\begin{equation}\label{eq:Ks}
    \begin{split}
        \text{K}_\ast(\hat{\bm{\tau}}) &= \begin{pmatrix}
            V(\bm{\theta}_{i}^\ast, \bm{\theta}_{j} ; \hat{\bm{\tau}}) \,\,\, Y(\bm{\theta}_{i}^\ast, \bm{\theta}_{j} ; \hat{\bm{\tau}}) \\
            Y^T(\bm{\theta}_{i}^\ast, \bm{\theta}_{j} ; \hat{\bm{\tau}}) \, W(\bm{\theta}_{i}^\ast, \bm{\theta}_{j} ; \hat{\bm{\tau}}) 
            \end{pmatrix}\\
        \text{K}_{\ast\ast}(\hat{\bm{\tau}}) &= \begin{pmatrix}
            V(\bm{\theta}_{i}^\ast, \bm{\theta}_{j}^\ast ; \hat{\bm{\tau}}) \,\,\, Y(\bm{\theta}_{i}^\ast, \bm{\theta}_{j}^\ast ; \hat{\bm{\tau}}) \\
            Y^T(\bm{\theta}_{i}^\ast, \bm{\theta}_{j}^\ast ; \hat{\bm{\tau}}) \, W(\bm{\theta}_{i}^\ast, \bm{\theta}_{j}^\ast ; \hat{\bm{\tau}}) 
        \end{pmatrix}.\\
    \end{split}
\end{equation}
We again note that during optimization, the hyperparameters controlling scales and amplitudes can explore all positive real numbers, while the mean function and the hyperparameter, $\Delta q_{BS}$ introduced in Eq.~\ref{eq:d_Y}, can explore all real numbers. In practice, we set the initial values of $\bm{\tau}$ to be random numbers close to zero and find that irrespective of initialization, we obtain the same optimized $\hat{\bm{\tau}}$. Similarly, we find that the hyperparameters of the optimizer, such as the learning rate, do not affect our results. When we perform optimizations with higher learning rates and lower iteration numbers, we obtain the same results as when using a lower learning rate and higher iteration number. Finally, the results appear stable across different optimizers. We explored both ADAM and SGD, finding similar optimized results between them. We use the SGD optimizer in this work, as we noticed it converges to a solution faster than ADAM.

To summarize the CARPoolGP strategy, we provide a step-by-step algorithm. 
\begin{enumerate}
    \item Define a set of parameters, $B$, at which to generate base sample quantities $\tilde{Q}$.
    \item Define a set, $S$, of parameter islands, $\theta_{S}$, in which to generate surrogate sample quantities, $\tilde{Q}^S$. Ensure that base-surrogate pairs have some level of correlation between them. 
    \item Define the noise kernels and associated hyperparameters for the base, surrogate, and combined quantities: $C_{ij}$, $D_{ij}$, $X_{ij}$.
    \item Maximize the likelihood function to obtain the optimal set of hyperparameters, $\hat{\bm{\tau}}$.
    \item Emulate using Eq.~\ref{eq:CPprediction} to find the predicted mean $ Q(\bm{\theta}_{i}^\ast)$, and its associated predictive covariance $\sigma^2(\bm{\theta}_{i}^\ast, \bm{\theta}_{j}^\ast)$ at new locations in parameter space.
\end{enumerate}

\subsection{Active learning approach}\label{sec:activelearning}

The above outline for CARPoolGP says nothing about the optimal locations in parameter space to draw base or surrogate samples. Instead, it describes a process for performing emulation given base and surrogate samples that were designed to have correlated noise. We show in Section~\ref{sec:Toy} that randomly distributing base samples and uniformly placing surrogate samples reduces the variance on some emulated quantity. However, one could imagine that drawing samples at specific regions in parameter space could provide more or less benefit in the pursuit of variance reduction. Indeed, there has been research on sampling optimal locations in parameter space using both Gaussian process regression and Bayesian optimization \citep{Leclercq-2018}. Here, we implement a simplified approach that fits within the framework developed in the previous sections.

This section explores a sampling method that picks base-parameter locations based on their predicted influence over the emulator's variance reduction. We develop an approach such that sample locations are not chosen \textit{a priori} but instead on the fly using CARPoolGP. This approach equitably treats the parameter space by buttressing regions predicted to have enhanced variance with extra samples and sparsely collecting data in regions that do not require any additional assistance. 

As we have already seen, given a conditioned covariance matrix, $\Sigma(\hat{\bm{\tau}})$ one can emulate some quantity at a new parameter point $\bm{\theta}_{i}^\ast$, obtaining the predicted quantity $ Q(\bm{\theta}_{i}^\ast)$ and its variance $\sigma^2(\bm{\theta}_{i}^\ast, \bm{\theta}_{j}^\ast)$ using Eq.~\ref{eq:CPprediction}. Furthermore, the predicted variance is merely a function of the optimal hyperparameters in $\Sigma(\hat{\bm{\tau}})$ and the locations in the parameter space used in conditioning, $\bm{\theta}_{i}$, and emulating, $\bm{\theta}_{i}^\ast$ (Eq.~\ref{eq:Ks}); in particular, the variance is not explicitly dependent on the value of the quantities. This means that by incorporating more parameter space locations into the base and surrogate samples vector, we can predict the change in predictive variance following Eq.~\ref{eq:CPprediction} even before actually generating new samples at these parameter points. We use this change in predictive variance to explore a range of potential new parameter locations, called \textit{Candidate points}, finding which candidate provides the largest reduction in predictive variance across the entire parameter space. 

More rigorously, we consider a range of candidate points in the parameter space $K\equiv\{\theta_{i}|i=1, 2, ..., N_k\}$, $K \notin B$, $K\notin S$ and test points $T\equiv\{\theta_{i}| i=1, 2, ..., N_T\}$, $T \notin B$, $T\notin S$, $T\notin K$. Each candidate is individually incorporated into the base parameter vector, $\bm{\theta}_{i}$, and an associated surrogate with correlated sample variance is placed at a parameter island following some algorithm. In this work, we place surrogates on the nearest-neighbor islands, but other choices could also be made. These updated covariance kernels are used to predict the variance across the full set of test points using Eq.~\ref{eq:CPprediction}. From this we obtain the variance, $\sigma_k^2(\bm{\theta}_{i}^\ast, \bm{\theta}_{j}^\ast)$ from points in the test set, $T$. We place $k$ in the subscript of $\sigma^2$ to make clear that the variance is unique for each candidate point $\theta_{k}$. We summarize the effect of adding the candidate as the trace of the predictive covariance matrix, which we call $\beta_k$, 
\begin{equation}\label{eq:betaeq}
    \beta_k = \text{Tr}\left(\sigma_k^2(\bm{\theta}_{i}^\ast, \bm{\theta}_{j}^\ast)\right)= \sum_{t=1}^{t=N_T}  \sigma_k^2(\bm{\theta}_{t}', \bm{\theta}_{t}').
\end{equation}
The candidate point that obtains the smallest value for $\beta_k$ is considered the best place to perform the next sampling, $\hat{\theta}_{k}$, as it will provide the largest variance reduction when emulating across the entire space.

This process is not limited to finding a single next-parameter location, $\hat{\theta}_{k}$. Instead, a batch of future points can be obtained by repeating the above process and updating the covariance matrices at each iteration. In this way, one can choose an arbitrary number of points to sample next. However, it should be noted that the parameters chosen are done with respect to the current best-fit set of $\hat{\bm{\tau}}$. It could be that $\hat{\bm{\tau}}$ is sensitive to the addition of numerous new parameters, so being frugal in the number of new parameter points or testing the sensitivity of $\hat{\bm{\tau}}$ with respect to the number of additional parameter points is important. 

There are three further generalizations to this approach, which we briefly identify to assist in better understanding this process. However, we do not explore them further in this work. First, in Eq.~\ref{eq:betaeq}, we only compute the predictive variance of one quantity, but this is a simplifying choice that can be made arbitrarily more complex. In practice, one can choose to predict a plethora of quantities and sum their individual predictive variance traces following some normalization to remove any units. The quantity with the most dominant normalized variance will likely determine the choice of successful candidate points, but for those interested in a variety of quantities drawn from an individual sample, using this more generalized approach may be desirable. Second, the choice to use the trace in computing Eq.~\ref{eq:betaeq} is somewhat arbitrary, and in general, other compression operations can be used, such as the determinant. In practice, we use the trace, as this is typically a computationally cheap operation and, more importantly, the most straightforward intuitive measure of the hyper-parameter space's predictive covariance. Finally, the abovementioned process focuses on placing new coordinates based on base samples, with surrogates located at the nearest neighbor parameter island. However, this process could be extended to remove any association with the candidate point. Then, the active learning process could determine whether the candidate point should be placed as a base or as a surrogate at a new parameter island location. In this second case, one must find the best base to correlate with the new surrogate.

\subsection{A toy example}\label{sec:Toy}
\begin{figure*}[t]
    \centering
    \includegraphics[width=\textwidth]{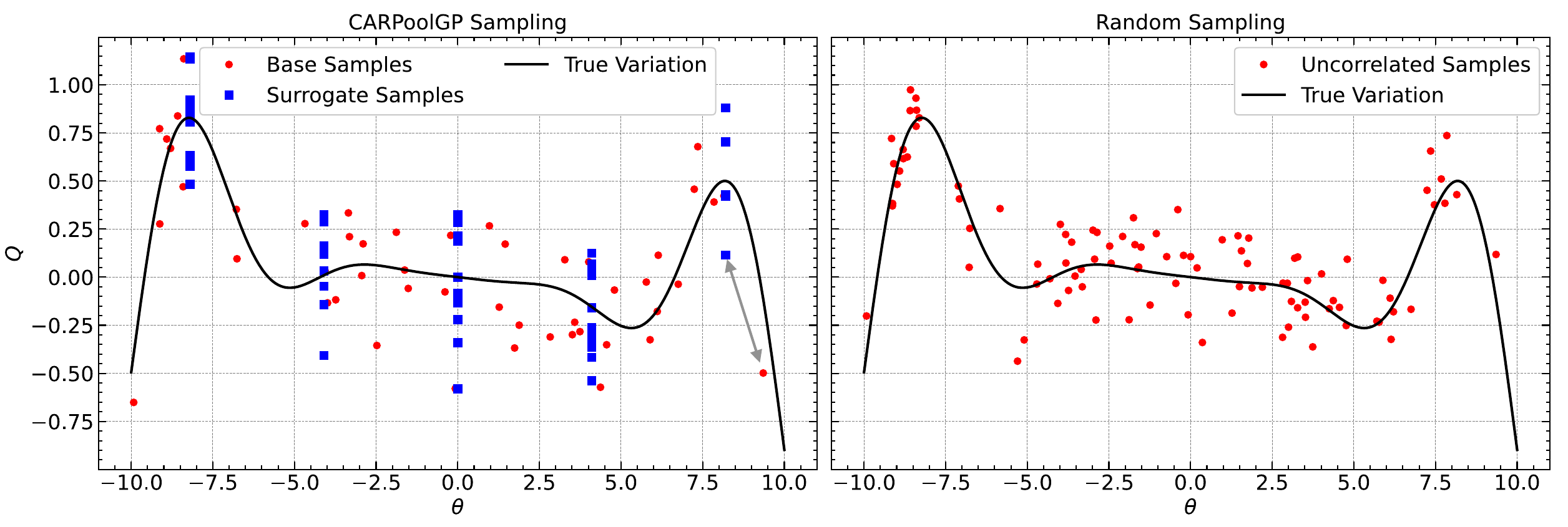}
    \caption{The sampling for our toy model, which we use to explore the efficacy of CARPoolGP. The left panel shows fifty base samples in red and fifty surrogate samples in blue, where each base sample has a sample variance correlated to a surrogate sample at the nearest parameter island. The grey double-headed arrow indicates a pair of samples that have been correlated with one another. In the right panel, we show an alternative sampling approach in which one hundred data points have been drawn from a uniform random distribution but contain uncorrelated sample variance. Both panels show in black the true variation of the quantity $Q$, which follows Eq.~\ref{eq:ToyModel}.}
    \label{fig:toysampling}
\end{figure*}

We now illustrate the potential of CARPoolGP and the active learning approach on a simple one-dimensional toy model where the base and surrogate samples are generated via the same process. We consider some mean quantity, $Q$, which has a functional form, 
\begin{equation}\label{eq:ToyModel}
\begin{split}
     Q(\theta_{i}) &= a \theta_{i} + b\theta_{i}^3\sin(\theta_{i})\\
    \tilde{Q}(\theta_{i}) &=  Q(\theta_{i})  + \epsilon,
\end{split}
\end{equation}
where $a$ and $b$ are some constants, and $\theta_i$ is our independent, one-dimensional parameter. Once again, this quantity and its functional form are completely arbitrary and for the sake of a better understanding of how CARPoolGP works. As we will show in later sections, higher dimensional parameter spaces with more complicated processes work just as well with CARPoolGP. We will follow the procedure outlined at the end of Section~\ref{sec:CPTheory}.

\subsubsection{Generate random parameters and base sample quantities}

We build the base quantities by sampling a uniformly random distribution within the domain $[-10, 10]$ to generate a set of parameter points $B\equiv\{{\bf \theta}_{i}| i=1, 2, ..., N\}$, and quantities, $\tilde{Q}({\theta}_i)$. We choose $N=50$ and draw $\epsilon$ from a normal distribution with $\langle\epsilon\rangle = 0$ and $\langle\epsilon^2\rangle = \sigma_Q^2$ which we set to a constant value of $\sigma_Q = 0.1$. 

\subsubsection{Generate parameter islands at which to extract surrogate sample quantities}

We then generate parameter islands in the set $S\equiv\{\theta_{i}| i=1, 2, ..., N_S\}$ by linearly spacing $N_S=5$ points in the range $[-8, 8]$ with the same process as defined in Eq.~\ref{eq:ToyModel}. For each base sample, the island closest to the parameter is identified, and a surrogate sample is drawn at this island location, $\theta_i$, to generate $\tilde{Q}^S(\theta_i)$ where the noise, $\epsilon_s$, is perfectly correlated with the noise of the base simulation (i.e., the same amplitude of the noise is used $\epsilon_{s, i} = \epsilon_{i}$).

The left panel in Fig.~\ref{fig:toysampling} shows the data we construct for this toy model. The red points are the base samples, each of which is not correlated with all of the others and lives at an independent point in the parameter space. The blue squares are surrogate samples that live together on the parameter islands. Each surrogate has correlated noise to a neighbor base sample. We show a two-headed arrow in grey to represent a base-surrogate correlation pair, but each base sample on the left panel is correlated with a surrogate on a parameter island, making a unique base-surrogate pair. The quantity, $ Q (\theta_i^\ast)$, has some true evolution at linearly spaced test locations within the domain [$-10, 10$]: $T\equiv\{\theta_{i}^\ast| i=1, 2, ..., N_T\}$ that follows Eq.~\ref{eq:ToyModel}. We set $N_T=1000$, and plot $ Q (\theta_i^\ast)$ in black.

The right panel of Fig.~\ref{fig:toysampling} shows a more typical approach, where samples are randomly drawn from a uniform distribution throughout the parameter space and are not correlated with one another. We will be interested in comparing CARPoolGP, with fifty base and fifty surrogate samples, to a standard GP that uses the purely random sampling approach with one hundred uncorrelated samples.

\subsubsection{Determine noise kernels}
\begin{figure*}[t]
    \centering
    \includegraphics[width=\textwidth]{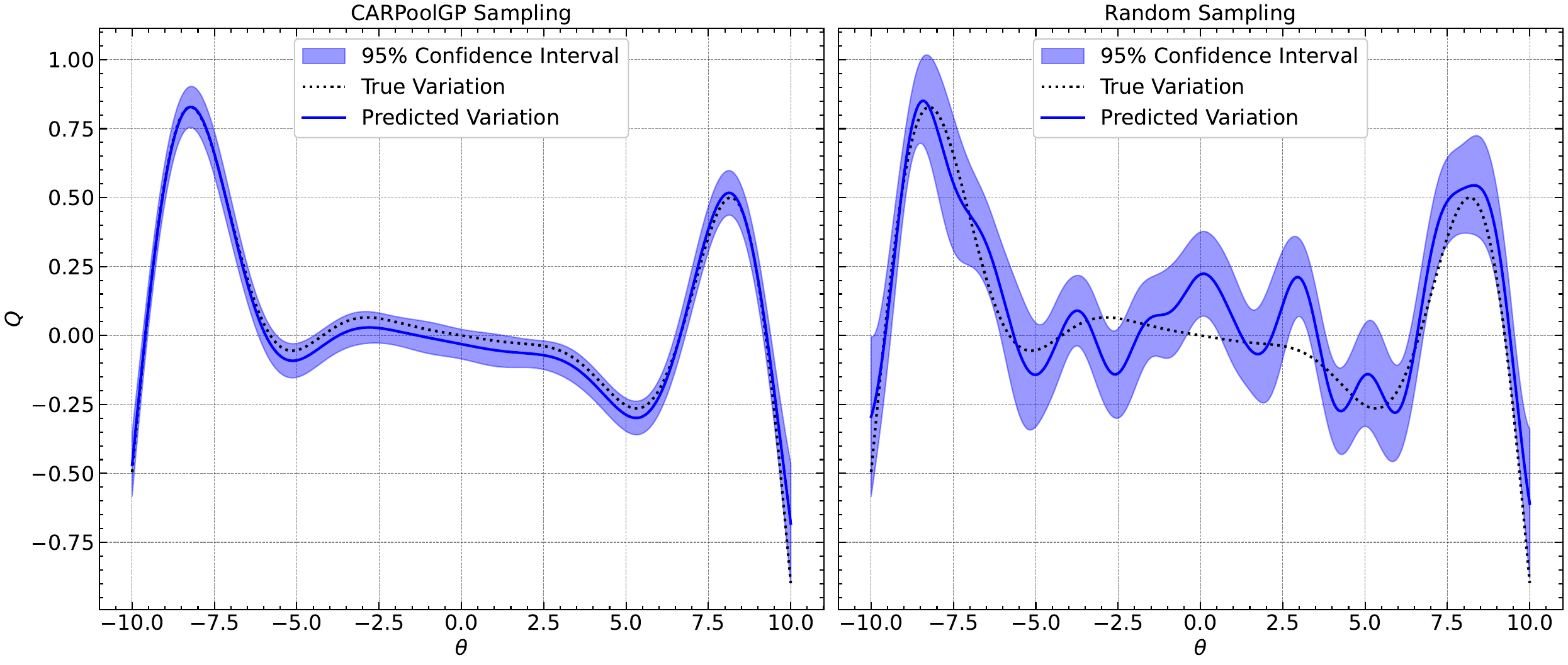}
    \caption{The CARPoolGP emulation with 95\% confidence intervals (left panel as a blue line and shaded region) and a standard sampling approach (right panel as a blue line and shaded region). In both subplots, we show the true quantity variation as a black dotted line. Both emulations have the same cost in terms of the number of samples, but when using the CARPoolGP approach -- with correlated base-surrogates samples that live on parameter islands, as shown in Fig. \ref{fig:toysampling} -- we obtain a much more accurate emulator with a greatly reduced predictive variance. Unlike the following figure (Fig.~\ref{fig:toyactivelearning}), in this figure, no active learning has been used to select the sample locations}
    \label{fig:toyfits}
\end{figure*}
For both base and surrogate samples, we use a radial basis function defined in Eq.~\ref{eq:rbf} to describe the smooth varying component of the covariance. Base and surrogates are drawn from the same underlying process and with the same level of sample variance, so the hyperparameters, $\bm{\tau}$, are shared across both matrices. 
\begin{equation}
\begin{split}
    V_{ij} = \alpha\exp\left(-\gamma \, d_E(\theta_{i} - \theta_{j})^2\right)\\
    W_{ij} = \alpha\exp\left(-\gamma \, d_E(\theta_{i} - \theta_{j})^2\right).
\end{split}
\end{equation}
The only difference between the two matrices is the parameters that are used to generate them. $V$, uses the base samples, while $W$ uses the surrogate samples. The full covariance for the base samples and the surrogate samples can be written following Eq.~\ref{eq:cov_C} and Eq.~\ref{eq:cov_D}
\begin{equation}
\begin{split}\label{eq:toyCD}
    C_{ij} &= \alpha\exp\left(-\gamma \, d_E(\theta_{i} - \theta_{j})^2\right) + \sigma_Q^2\mathcal{I}\\
    D_{ij} &= \alpha\exp\left(-\gamma \, d_E(\theta_{i} - \theta_{j})^2\right) + \sigma_Q^2\mathcal{I}.
\end{split}
\end{equation}
We choose the kernel that describes the smooth covariance between the base and surrogate samples to be an RBF following Eq.~\ref{eq:cov_X}, but we set the additional parameter, $\Delta q_{BS} =0$, as the processes between the base and surrogates are the same. We use the same scale and amplitude parameters for the $V_{ij}$ and $W_{ij}$ matrices to define the covariance between base and surrogate samples,
\begin{equation}
    Y_{ij} = \alpha\exp\left(-\gamma \, \left(d_E(\theta_{i}, \theta_{j})^2\right)\right).\\    
\end{equation}
To relate the base samples to the surrogates, we use the fact that we have set a perfect correlation between the sample fluctuations and, therefore, set the $M$ matrix to
\begin{equation}
    M_{ij} = \sigma_Q^2\delta_{ij},
\end{equation}
where the $\delta_{ij}$ is a delta function that is $1$ at locations of base-surrogate pairs, and $0$ elsewhere. Recall that the distance between parameter space locations in $Y_{ij}$ and $M_{ij}$ are evaluated between base and surrogate samples.
Following Eq.~\ref{eq:cov_X}, we then have
\begin{equation}\label{eq:toyX}
\begin{split}
    X_{ij} = &\alpha\exp\left(-\gamma \, \left(d_E(\theta_{i} - \theta_{j})^2\right)\right) + \sigma_Q^2.\\
\end{split}
\end{equation}
We can now build the block covariance matrix containing all of these components following Eq.~\ref{eq:sigma} where $\bm{\tau}$ is the vector of hyperparameters, $\bm{\tau}=(\alpha, \gamma, \sigma^2_Q)$

Because we are interested in studying the effect of using correlated samples, we build the same covariance matrices for the scenario where we have no surrogate simulations, and each sample is simply uncorrelated (right panel of Fig.~\ref{fig:toysampling}.) In the scenario of completely uncorrelated samples, the block covariance matrix is equivalent to a standard RBF function with three free hyperparameters describing the amplitude and scale of the RBF and a diagonal term representing the variance level of the samples. 

\subsubsection{Maximize the likelihood function}

We use the Gaussian likelihood function as defined in Eq.~\ref{eq:CPliklihood} and choose uninformative priors for $\mu_B$ and $\mu_S$, but allow them to be learned as additional hyperparameters in the regression. We have written CARPoolGP in the \texttt{JAX}\footnote{\url{jax.readthedocs.io}} \citep{jax2018} programming language, allowing for auto-differentiation support and efficient gradient descent optimization. We minimize the negative log of the likelihood function to obtain an optimal set of hyperparameters, $\hat{\bm{\tau}}$ using Stochastic Gradient Descent (SGD) from the \texttt{optax}\footnote{\url{optax.readthedocs.i}} package \citep{optax-2020}. To compare the effect of CARPoolGP, we perform an identical optimization process on a set of one hundred uncorrelated samples.

\subsubsection{Emulate the quantity}
\begin{figure*}[t]
    \centering
    \includegraphics[width=\textwidth]{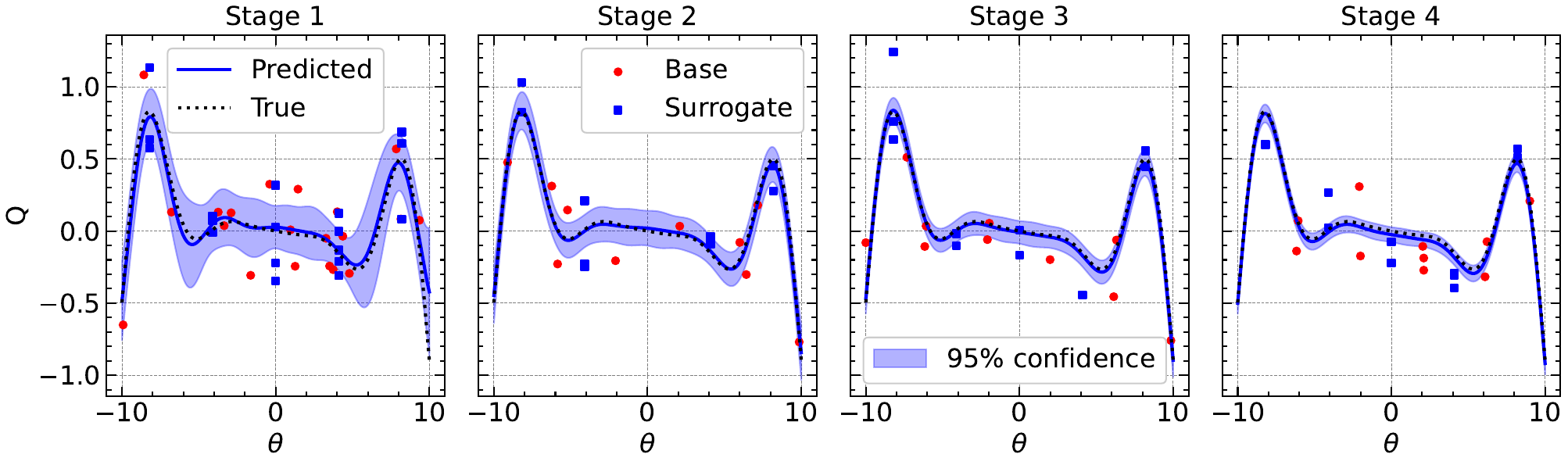}
    \caption{A demonstration of the active learning approach to CARPoolGP. Each panel represents a stage in the active learning process, and the first panel shows a standard CARPoolGP approach with twenty base and twenty surrogate samples. In the next subplot, we show the new ten base points (red dots) and ten surrogate points (blue) chosen for sampling at the next stage. The results of emulation with this parameter sampling arrangement are shown in the 95\% confidence interval. The next two panels show the following two stages of active learning, the last containing a cumulative total of fifty base samples with fifty associated surrogates as in the left panels of Fig.~\ref{fig:toysampling} and Fig.~\ref{fig:toyfits}.}
    \label{fig:toyactivelearning}
\end{figure*}

We now have all we need to perform an emulation at sample points from the set $T$ using Eq.~\ref{eq:CPprediction}. In Figure~\ref{fig:toyfits}, we show the results of this sampling, fitting, and emulation procedure. The left panel shows the results using the CARPoolGP method, where the grey line represents the emulated quantity, $ Q(\bm{\theta}_i^\ast)$, and the shaded region shows the 95\% confidence region evaluated from $\sigma^2(\bm{\theta}_i^\ast, \bm{\theta}_i^\ast)$. We compare this with the right panel that contains the prediction and error when the samples are uncorrelated and drawn from a uniform distribution in the parameter space. Both panels were generated with the same `cost,' meaning that they both contain one hundred samples. However, the variance in predicted quantities from the CARPoolGP panel is much smaller than in the typical case and contains fewer spurious fluctuations.

\subsubsection{Apply active learning}

We now show that by performing our sampling and emulation in stages, we can choose the next places to sample that will provide the largest variance reduction in our final emulation. We follow the procedure outlined in Section~\ref{sec:activelearning} and test the efficacy of the active learning approach against the CARPoolGP approach without active learning from the left panel of Fig.~\ref{fig:toyfits}.

We follow the outline above for the CARPoolGP case but set $N=20$. The emulator is trained with the same kernels as previously described but now with only twenty randomly sampled points and their associated surrogates, which live amongst the same five-parameter islands as the above case. The base and surrogate samples and the CARPoolGP prediction from this first stage are shown in the left-most panel of Fig.~\ref{fig:toyactivelearning}. This sparse sampling of points leads to a large predictive variance, particularly at undersampled locations (e.g., $\theta=-5)$.

\begin{figure}[h]
    \centering
    \includegraphics[width=0.49\textwidth]{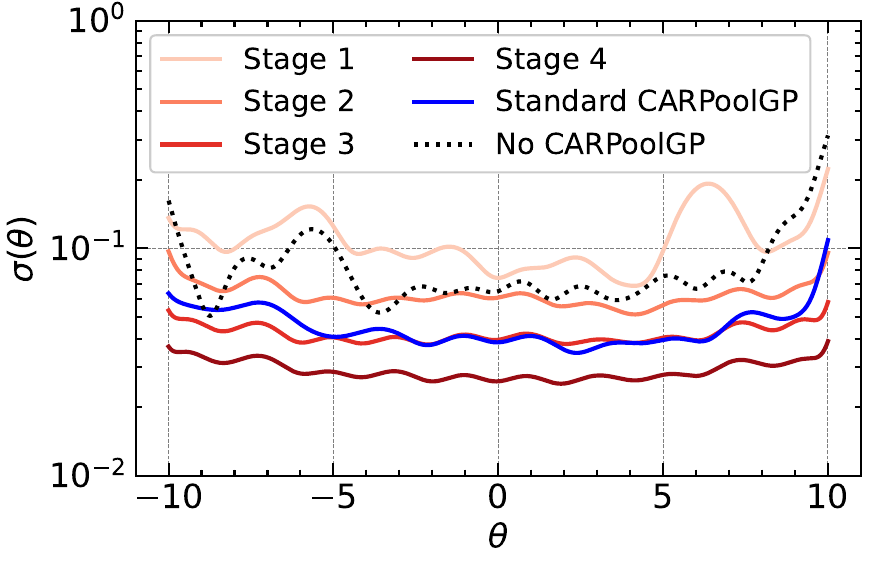}
    \caption{The standard error $\sigma(\theta)$ is shown as a function of test parameter space locations, $\theta^\ast$, for the active learning stages (increasing dark red lines), the standard CARPoolGP approach (blue line), and No CARPoolGP (dashed black line). The first stage of active learning contains 20 base and 20 surrogate samples, and each subsequent stage adds 10 base and 10 surrogate samples following the active learning method, such that stage 4 has a total of 50 base and 50 surrogate samples, as does the case of standard CARPool, while the no-CARPoolGP case has the same total of 100 samples. CARPoolGP vastly outperforms the random sampling approach and is further boosted when combined with active learning. }
    \label{fig:toyvarreduc}
\end{figure}

Following Section~\ref{sec:activelearning}, we first choose the number of new samples we want to generate, $N_t$, and a set of candidate points to pick from, $N_{c}$. To be clear, each new parameter location, $\theta_j$, added is chosen from a unique set of candidate points $K_{c, j}\equiv\{\theta_{i}|i=1, 2, ..., N_c\}$, $K_{c, j} \notin B$, $K_{c, j}\notin S$, $K_{c, j} \notin T$. The nearest neighbor parameter island is identified for each candidate point, and an associated surrogate sample is placed on this island, increasing the population by one.

We set $N_t=10$, and for each point, we check $N_c=512$ candidates and associated surrogates within the domain of the problem. Now we iteratively add each candidate point and associated surrogate to the covariance matrices to recalculate $\Sigma(\hat{\bm{\tau}})$, $K_\ast(\hat{\bm{\tau}})$ and $K_{\ast\ast}(\hat{\bm{\tau}})$, and compute the error on the set of test points following Eq.~\ref{eq:CPprediction}. We then compute the trace of the predictive covariance matrix as in Eq.~\ref{eq:betaeq} and save this value into the vector, $\bm{\beta}_c$. We find the best next place to sample the space as:
\begin{equation}
    \bm{\theta}_{j} = K_{c,j}[\text{argmin}(\bm{\beta}_c)].
\end{equation}

Repeating this process $N_t$ times provides a set of new locations to perform base sampling. Using the model of Eq.~\ref{eq:ToyModel}, we compute the value of $\tilde{Q}$ at the new base and surrogate parameter locations and retrain the CARPoolGP emulator to obtain a new set of $\hat{\bm{\tau}}$. In the second panel of Fig.~\ref{fig:toyactivelearning}, we show the results of our second stage of active learning. The ten new base samples are placed as red points, and the associated surrogates as blue squares. Not surprisingly, we find that the regions of the highest variance in the first subpanel are the regions that CARPoolGP's active learning procedure targets to sample next (e.g., $\theta=-5$). From this addition, we further find that the accuracy of the mean prediction is enhanced, and the variance is vastly reduced.

As shown in the third and fourth panels, we performed two more active learning stages. Each shows the next base samples that were chosen and their associated surrogates. Following the final step, our parameter space has been sampled one hundred times, with fifty base samples and fifty surrogates. We wish to compare three errors: CARPoolGP vs. No CARPool, CARPoolGP vs. CARPoolGP active learning, and CARPoolGP active learning vs. No CARPool. These comparisons are presented in Fig.~\ref{fig:toyvarreduc}, where the error, $\sigma(\theta^\ast_{ii}) = \sqrt{\sigma^2(\theta^\ast_i, \theta^\ast_i)}$, is shown as a function of the parameter space location. The increasingly dark red lines represent the active learning stages, starting with twenty base and twenty surrogate samples (lightest red) and ending with fifty base and fifty surrogate samples (darkest red). The scenario in which CARPoolGP is used with no active learning but contains fifty bases and fifty surrogates is shown in blue. We finally show the error from typical random sampling with one hundred uncorrelated samples in dashed black.

Here, we see that by performing active learning, we get a $\sim 36\%$ reduction in $\sigma(\theta^\ast)$ compared to the regular CARPoolGP, and $\sim 65\%$ better than the standard random sampling approach. Even if active learning is not used, CARPoolGP outperforms the error in random sampling by $\sim 45\%$.  To put these results in terms of costs, assuming that the error is reduced by $1/\sqrt{N}$, Fig.~\ref{fig:toyvarreduc} says that one would require $\sim 3$ times more samples to achieve the same error as if one did not use CARPoolGP, and $\sim 8$ times more samples to achieve the error in the active learning case. It should be noted that here, and in the rest of this work, we only use active learning to reduce the variance on a single quantity, which may not guarantee the same variance reduction on quantities not explicitly used in the active learning procedure. In general, more quantities can be used, and further research is required to investigate the changes in variance across quantities. We leave this multi-quantity active learning approach as an interesting avenue for further research.

\section{Simulations}\label{sec:simulations}
The following subsections detail the generation of our simulation suite and the context in which CARPoolGP is used. We begin with an overview of the galaxy formation model used in this work, and the parameter space spanned in the simulation suite. We then discuss the generation of base and surrogate simulations and how we utilize CARPoolGP active learning to iteratively choose future parameter space coordinates for simulations.

\subsection{Simulation setup}\label{sec:sim_setup}
The core CAMELS suite \citep{Villaescusa-Navarro-2021} contains (at the time of writing this paper) $5,516$ cosmological hydrodynamical simulations and $5,164$ of their associated N-body counterparts, with comoving volumes of $(25\, \Mpc\, h^{-1})^3$. CAMELS spans a range of galaxy formation models (SIMBA \citep{Dave-2019}, IllustrisTNG \citep{Weinberger-2017, Pillepich-2018}, Astrid \citep{Bird-2022, Ni-2022}), Magneticum \citep{Dolag-2015}, Eagle \citep{Crain-2015}, Enzo \citep{Bryan-2014}, Ramses \citep{Teyssier-2002}, and their respective astrophysical and cosmological parameter spaces. In this work, we only consider IllustrisTNG's galaxy formation model, which is built atop Illustris \citep{Genel-2014, Vogelsberger-2014} and takes advantage of the AREPO Tree-PM moving mesh code \citep{Springel-2010}, kinetic galactic winds and feedback following \cite{Springel-Hernquest-2003}, and black hole feedback with three modes: thermal, kinetic, and radiative \citep{Pillepich-2018}. 

\defcitealias{Ni-2023}{Ni23}

The flagship CAMELS suite parameterizes IllustrisTNG's galaxy formation model with four astrophysical parameters corresponding to supernova and AGN feedback, but more recently \cite{Ni-2023} (henceforth \citetalias{Ni-2023}) extended this set to include 2048 simulations sampled from a Sobol sequence \citep{Sobol-1967} over a 28 dimensional astrophysical and cosmological parameter space in the IllustrisTNG model (TNG-SB28). 

This work explores the same 28-dimensional parameter space as in \citetalias{Ni-2023}. The parameter names, a brief description, and the bounds can be found in Appendix A of \citetalias{Ni-2023}. We also add a parameter to control the desired halo mass of the main halo in the zoom-in region. We limit this mass to a range of $13 \leq \log( M_{200}/\left[M_\odot\,h^{-1}\right])  \leq 14.5$, where $M_{200}$ is the total enclosed mass inside a radius $R_{200}$ which describes the radius at which the density is two hundred times the critical density, $\rho_c$, of the universe. We consider this mass range as it is relevant to current X-ray and upcoming CMB experiments.

We further highlight that our sampling method differs from TNG-SB28 in that we do not sample all of our simulations from the same Sobol sequence, but instead, we utilize the active learning method outlined in Section~\ref{sec:activelearning}. We discuss this process and its implications in the following section.

At each parameter space location, $\bm{\theta}_i$, we run three separate simulations: A parent simulation used to identify a halo for zooming in, a hydrodynamical zoom-in, and dark matter-only analog zoom-in of the halo chosen from the parent simulation. We first create a $(200 \Mpc\,h^{-1})^3$ \textit{parent} box with $256^3$ dark matter particles that have mass $M_{dm} = 1.323\times 10^{11}\Omega_m \,M_\odot\,h^{-1}$ where $\Omega_m$ is the matter density of the universe and is varied in the 28-dimensional parameter space. Note that this is a very low-resolution simulation and exists only for the purpose of generating a large cosmological box with a large population of different halo masses. Initial fluctuations are generated at $z=127$ using MUltiScaleInitialConditions (MUSIC) \citep{Hahn-Abel-2011} with second-order Lagrangian Perturbation Theory, and evolved to $z=0$ using AREPO \citep{Weinberger-2020}. We identify halos in the parent box using the Friends of Friends algorithm \citep{Davis-1985}. We then look for a halo at $z=0$ with mass $M_{200}$ (which is calculated by SUBFIND \citep{Springel-2001} around the particle with the minimum energy and with respect to the critical density) closest to the desired mass $\theta_{i, \rm Mass}$. The true mass of the halo is used to replace the mass direction in this vector, $\theta_{i, \rm Mass} = M_{200}$.

All particles within $6\times R_{200}$ of the chosen halo are identified and traced back to their locations in the initial conditions, where a bounding rectangular Lagrangian region is drawn to encapsulate all of the traced particles. We chose such a large region surrounding each halo to ensure reduced low-resolution dark matter contamination in the final zoom-in simulation \citep{Onorbe-2014}. The hydrodynamical zoom-in simulation is then run centered on the chosen halo, where the gas particles have $M_{\rm gas} = 1.26\times 10^{7}\left(\Omega_b/0.049\right) \,h^{-1} M_\odot$ and the high-resolution Lagrangian region contains dark matter particles with mass $M_{\rm DM, High Res} = 6.49\times 10^{7}\left((\Omega_m - \Omega_b)/0.251\right) \,h^{-1} M_\odot$, matching the resolution of the $(25 \Mpc\,h^{-1})^3$ CAMELS boxes. For completeness, we perform associated dark matter-only zoom-in analogs to match the above using the same method and with the same resolution.

\subsection{Base-surrogate simulation pairs}\label{sec:BS_pairs}

The benefit of CARPoolGP described in Section~\ref{sec:CPTheory} relies on correlations in the sample fluctuations between the bases and surrogates. In the toy model (Section~\ref{sec:Toy}), we forced a perfect correlation by adding the same Gaussian distributed noise realization to base-surrogate pairs, but in the situation where the quantity of interest is extracted from a simulation, the sample variance, often called ``cosmic variance", derives from the random seed describing the phases and amplitudes of initial fluctuations in the simulations themselves. The base and surrogate simulations are then built by performing a zoom-in of the same halo at two separate points in parameter space - one at a unique parameter space location (base), and one on a parameter island (surrogate). 

The process for creating this pair starts with the generation of the base parent box and choosing a halo at some parameter space location, following the direction of the previous section. The surrogate halo is now chosen by first finding the closest (smallest Euclidean distance) parameter island, $\bm{\theta}_S$, to the parameter space coordinate of the base simulation. We then generate a parent box for the surrogate following the same process as the base, but at the parameter location, $\bm{\theta}_S$. While the amplitudes of the MUSIC initial conditions depend on cosmological parameters, the random numbers controlling the phases of the initial fluctuations of the surrogate simulation are matched to the base simulations so that their realizations are only affected by the changes in parameter space locations and have correlated sample variance. We determine the halo catalog in the surrogate simulation using FoF, and the surrogate halo is then found by performing a bijective matching between the base halo and the surrogate parent box to find the most probable surrogate halo match. Just as in base simulations, the mass parameter $\theta_{S, \rm Mass}$ is set to the true value of the surrogate halo mass: $\theta_{S, \rm Mass} = M_{200}$ where $M_{200}$ is computed through SUBFIND. We note that the differences in cosmology translate to differences in mass between a base-surrogate pair. We describe this effect on the distribution of masses in the full suite of simulations in Sec.~\ref{sec:AL_steps}. Now that both the base and surrogate have their associated parent boxes and zoom-in regions, we run the hydrodynamical and dark matter-only analogs following the previous section.

\subsection{Sampling stages}\label{sec:sampling_stages}

In this work, we apply the active learning procedure of Section~\ref{sec:activelearning}, unlike the Sobol sequence used in \citetalias{Ni-2023}, or the Latin hypercube in \citet{Villaescusa-Navarro-2021}. Our goal is to perform simulations at locations in the 29-dimensional parameter space (28 IllustrisTNG parameter dimensions and 1 mass parameter dimension sampled logarithmically) that optimally reduce the predictive covariance of some chosen quantity, as we presented in the toy model of Section~\ref{sec:Toy}. 

This process requires the choice of 1.) some quantity to extract from the simulations, 2.) a covariance kernel trained on a set of existing simulations, 3.) candidate parameter space locations to consider running future simulations, and 4.) an independent set of test parameter space locations to compute the variance over. We address each of these steps in the following paragraphs.

We focus our active learning on the Compton $Y$ parameter defined as, 
\begin{equation}\label{eq:y200}
    Y_{200, c} = \dfrac{\sigma_T}{m_e c^2} \int_V P_e(\bm{r}) dV.
\end{equation}
Where, $\sigma_T$ is the Thompson cross section, $m_e$ is the mass of an electron, $c$ is the speed of light, and the integral is done over the electron pressure ($P_e$) within a sphere of radius $R_{200}$. We will refer to this quantity as $Y$ in the text for brevity. 

As discussed in Section~\ref{sec:intro}, future CMB and X-ray experiments will resolve the tSZ effect in halos down to the $M_{200} \sim 10^{13} M_\odot\,h^{-1}$ scale, providing constraints on astrophysical processes such as AGN feedback and galactic winds. In addition, $Y$ is a proxy for halo masses following commonly used self-similar relations and also represents the thermodynamic properties of the halo gas \citep{Nagai-2006, Battaglia-2012, Ettori-2013}. In numerous recent studies using CAMELS, the $Y$ signal has been explored for the purpose of finding improved scaling relations \citep{Wadekar-2023a, Wadekar-2023b}, constraining galaxy environments \citep{Hadzhiyska-2023a}, and predictions on feedback constraints from the CGM \citep{Moser-2022}. In these works using CAMELS, the halo mass is typically limited to $10^{13}\,M_\odot\,h^{-1}$, because of the small number of halos that exist across the CAMELS suite above this range. By including a reduced variance emulator of $Y$ for high-mass halos, we hope to provide a means for improving the mass coverage of these works across the full range of parameter space.

The first set of base simulations, $B_1\equiv\{\bm{\theta}_{i}| i=1, 2, ..., 128\}$, is drawn from a Sobol sequence with 128 parameter space locations across the 29-dimensional parameter space that spans the prior ranges shown in \citetalias{Ni-2023}. The parameter islands, $S\equiv\{\bm{\theta}_{i}| i=1, 2, ..., 128\}$, are chosen to be a set of one-hundred and twenty-eight parameter space coordinates from a Sobol sequence with initialization such that $S \notin B_1$, and the set of surrogates is chosen by finding the nearest neighbor island to each base simulation. The parameter space islands used in this first stage are kept constant throughout all the active learning stages, and each surrogate is always chosen in this way. We run \textit{Stage 1} of the group-scale zoom-in simulation suite with the first set of 128 base and 128 surrogate parameter space locations following Section~\ref{sec:sim_setup} and compute the resulting $Y$ from Eq.~\ref{eq:y200} for each simulation. 

Covariance kernels are generated following Section~\ref{sec:CPTheory} and using RBF kernels (Eq.~\ref{eq:rbf}) for smooth covariance matrices ($V$ in Eq.~\ref{eq:cov_C}, $W$ in Eq.~\ref{eq:cov_D}, $Y$ in Eq.~\ref{eq:cov_X}), and linear exponential kernels to correlate the sample variance ($M$ in Eq.~\ref{eq:cov_X}). Just as in the toy example of Section~\ref{sec:Toy}, the base and surrogate simulations are drawn from the same process (i.e., resolution), so the kernels can be written as:
\begin{equation}\label{eq:CARPoolGPKernel}
\begin{split}
        \bm{V}=\bm{W}=\bm{Y}&=\alpha \prod_{p=1}^{29} \exp\left(-\gamma_p \, d_E(\bm{\theta}_{i} - \bm{\theta}_{j})^2\right)\\
        \bm{M} &=\sigma_Q^2 \prod_{p=1}^{29} \exp\left(-\gamma_p \, d_E(\bm{\theta}_{i} - \bm{\theta}_{j})\right)\delta_{ij}.
\end{split}
\end{equation}
We minimize the negative log-likelihood function in Eq.~\ref{eq:liklihood} using an SGD optimizer and the extracted values of $Y$ from each halo in the simulation to obtain the optimal covariance kernel. We can now generate predictions and find the predictive covariance of Eq.~\ref{eq:CPprediction}. 

The next step in the active learning procedure is to test the location of candidate parameters and observe the relative effect on the predictive covariance matrix. To do this, we generate a new Sobol sequence with 1024 candidate points, $K\equiv\{\theta_{i}|i=1, 2, ..., 1024\}$, $K \notin B_1$, $K\notin S$ and another with test points $T\equiv\{\theta_{i}| i=1, 2, ..., 1024\}$, $T \notin B_1$, $T\notin S$, $T\notin K$. Each candidate point is given a surrogate at the nearest-parameter island. We iteratively compute the predictive covariance matrix at the test points after each candidate (and associated surrogate) is added, independently of all other candidates, to the predictive kernel. We then build the $\bm{\beta}$ vector following Section~\ref{sec:activelearning} by computing the trace of the predictive covariance matrix. The candidate and surrogate pair that produces the smallest $\beta$ value is saved, and these locations of the parameter space are chosen to perform the next base and surrogate simulations. We keep this base surrogate pair in the covariance matrix and repeat this process 128 times, with a new set of $K$ each time, until we have a new vector of 128 base and 128 surrogate parameter space locations to perform simulations at. This next set of 256 simulations is called \textit{stage 2}. 

We perform a third and fourth active learning step following the same process as in the above paragraph, except for a smaller number of simulations, 64 base and 64 surrogate. We call these stages \textit{stage 3} and \textit{stage 4}, and following this set, we are left with a total of 384 base and 384 surrogate simulations.

\begin{figure}[t]
    \centering
    \includegraphics[width=0.48\textwidth]{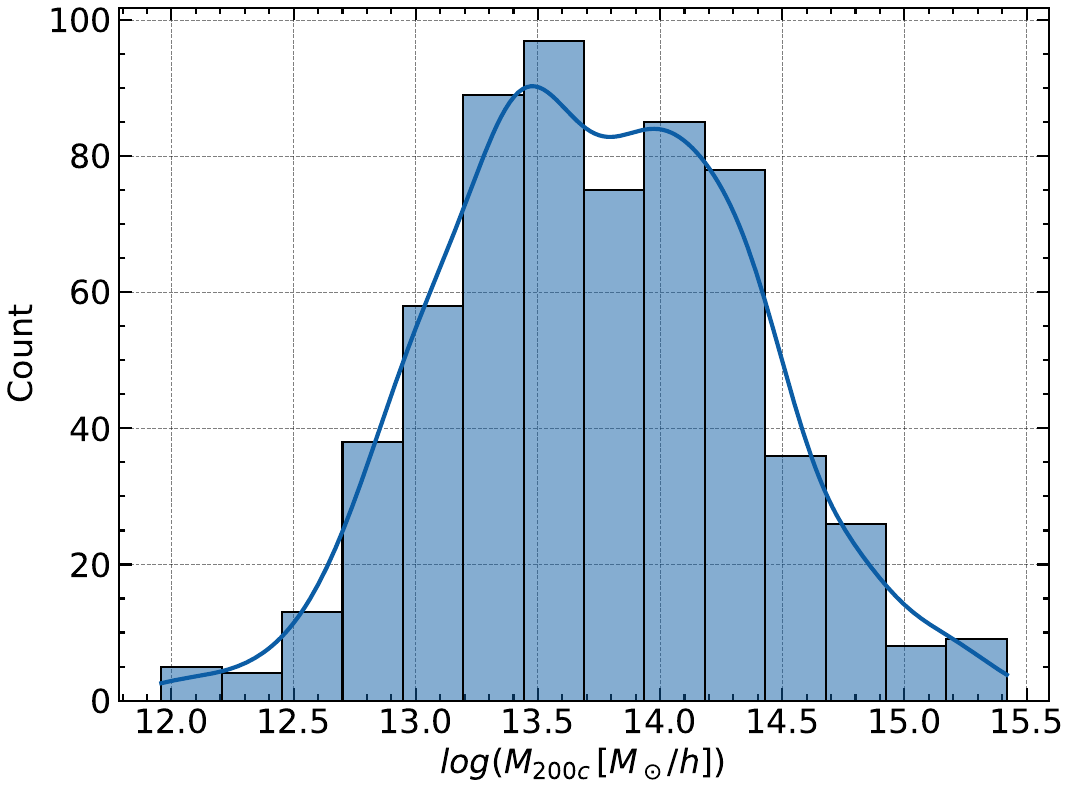}
    \caption{The distribution of masses from the suite of zoom-in simulations. We use a kernel density estimate of the distribution drawn as the blue line. For a Sobol sequence, the distribution in this space would be flat, i.e., have the same number of simulations in each bin, but it is clear that there are particular places in parameter space that are preferred using the active learning approach as opposed to a standard Sobol sequence.}
    \label{fig:mass_dist}
\end{figure}

Fig.~\ref{fig:mass_dist} shows the mass distribution produced through the full suite of simulations. There are two pieces of important information from this plot. First, the distribution of simulations in the parameter space is not uniform, as one would expect from a Sobol sequence, and instead contains regions of apparent preference. This clarifies that the resultant vector of parameters from the active learning CARPoolGP approach differs from what one would obtain using a Sobol sequence or Latin hypercube. The second is that the prior range of masses in $13\leq \log\left(M_\odot/\left[M_\odot\,h^{-1}\right]\right)\leq 14.5$ is extended due to the particular simulations. There are a few reasons for this. First, there are base simulations that have a desired mass parameter near the bounds of the prior range. In these scenarios, the halo with the most similar mass may be slightly outside of this range and cause the distribution to broaden. A more frequent occurrence is that a base simulation with a relatively high (low) extracted mass can contain a surrogate simulation at a parameter island with a larger (smaller) value of $\Omega_M$, causing the matching halo in the surrogate to be more massive (less massive) than the prior bounds. Both of these effects lead to a general broadening of the mass distribution. While we include all of the halos in the following analysis, we limit the emulated values to within the prior bounds and treat masses outside this range as extrapolations instead of interpolations. 

\subsection{Active learning stages}\label{sec:AL_steps}
\begin{figure}[t]
    \centering
    \includegraphics[width=0.48\textwidth]{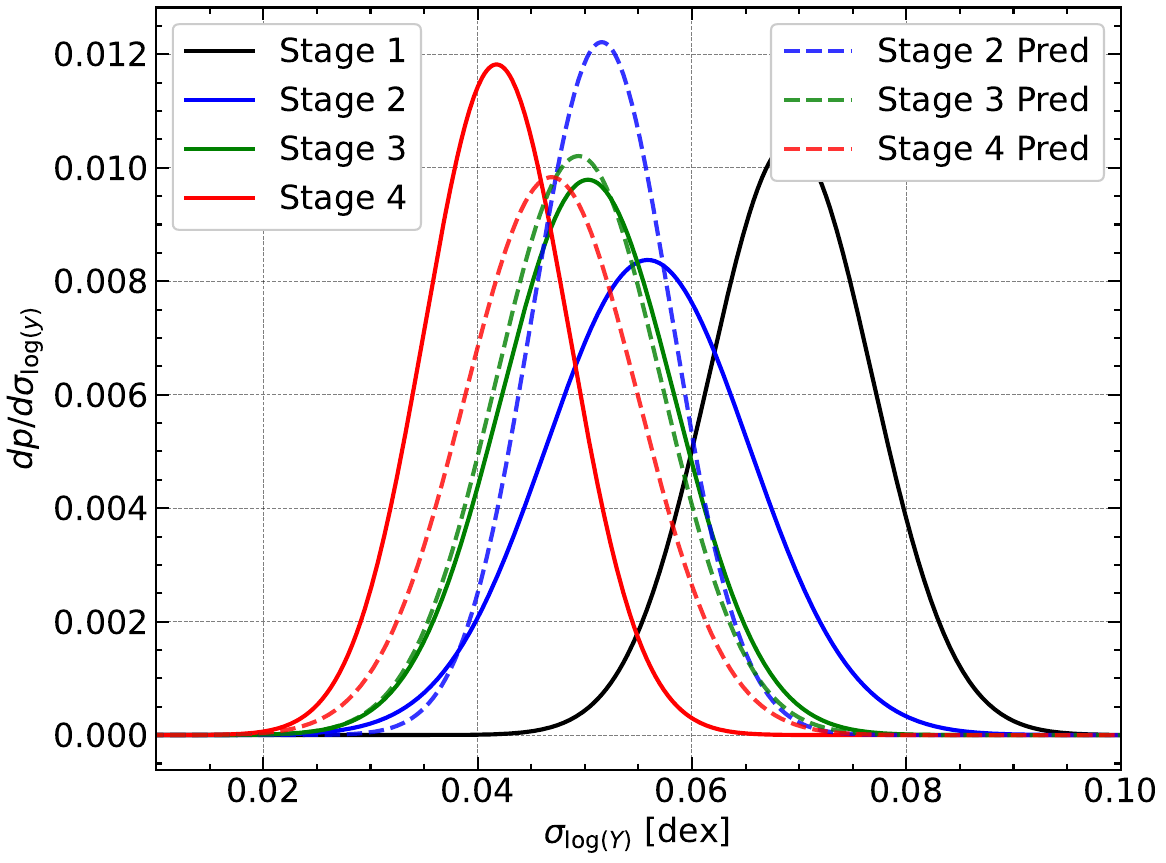}
    \caption{The predictive error is tested for each stage in the simulation suite over a set of $4096$ parameter space locations drawn from a Sobol Sequence. A Gaussian distribution is fit to the histogram of error values and is shown for each stage in the solid lines. We plot the expected histograms in dotted lines with their associated colors. Each stage reduces the mean of the predictive error distribution and decreases the width.}
    \label{fig:AL_hist}
\end{figure}

For each active learning step, we compute the predictive covariance of Eq.~\ref{eq:CPprediction} on a set of $4096$ test locations drawn from a sobol sequence within the priors described in \citetalias{Ni-2023} and the additional mass prior. We then fit a Gaussian to the distribution of the square root of the diagonal of the predictive covariance matrix, which represents the predictive standard deviation or uncertainty, $\sigma_Y$. In Fig.~\ref{fig:AL_hist}, we plot the fits for each stage in solid lines. As expected, there is a clear trend of decreasing predictive uncertainty with each stage. 

Considering the expected variance reduction from CARPoolGP prior to generating simulations is also interesting. We can do this by simply incorporating the suggested new points into the covariance matrices in Eq.~\ref{eq:CPprediction}. The dotted lines in Fig.~\ref{fig:AL_hist} represent what CARPoolGP predicts the distribution should look like after each step.

This experiment indicates that stage 2 did not improve the uncertainty as much as expected. On the other hand, stage 3 performed almost exactly as expected, and stage 4 outperformed expectations. This discrepancy could be due to several factors. First, the predictions of $Y$ using only the data from stage 1 contain some predictive uncertainty that could propagate into the expected results. An alternative reason could be that stage 2 doubled the number of simulations compared to stage 1. As noted in Section~\ref{sec:activelearning}, the active learning procedure operates under the assumption that the best-fit kernel parameters, $\bm{\hat{\tau}}$, slowly vary with respect to the addition of simulations. We used that assumption to incorporate many samples in stage 2, but we likely included too many in this stage, violating this assumption. After observing these results following stage 2, we reduced the number of new simulations in each stage by a half, leading to variance reduction at the predicted level.

The most conservative approach to active learning would be to add the simulations one at a time in any active learning step. This would ensure that the optimal hyperparameters stay very similar between added stages but would require running the simulations in serial, which would be infeasible for running hundreds of simulations in a timely manner. A balance needs to be met where the number of new simulations cannot exceed a level that significantly changes the values of the hyperparameters but also allows for a generous amount of simulations to be run simultaneously in parallel. We do not investigate this further here, leaving exploration of this balance to future work.

\section{Results and physical interpretations} \label{sec:results}

In this section, we present results using the zoom-in simulations and CARPoolGP. We demonstrate the new suite and emulator's utility in addressing questions in numerical galaxy formation and cosmology that were hitherto inaccessible. We begin by emulating the $Y-M$ relation at the fiducial IllustrisTNG parameter space location and with variations of individual parameters around it to study its dependencies. We then used CARPoolGP to emulate other halo summary statistics, allowing qualitative explanations of the observed parameter trends in the $Y-M$ relation. 

\begin{figure}[t]
    \centering
    \includegraphics[width=0.48\textwidth]{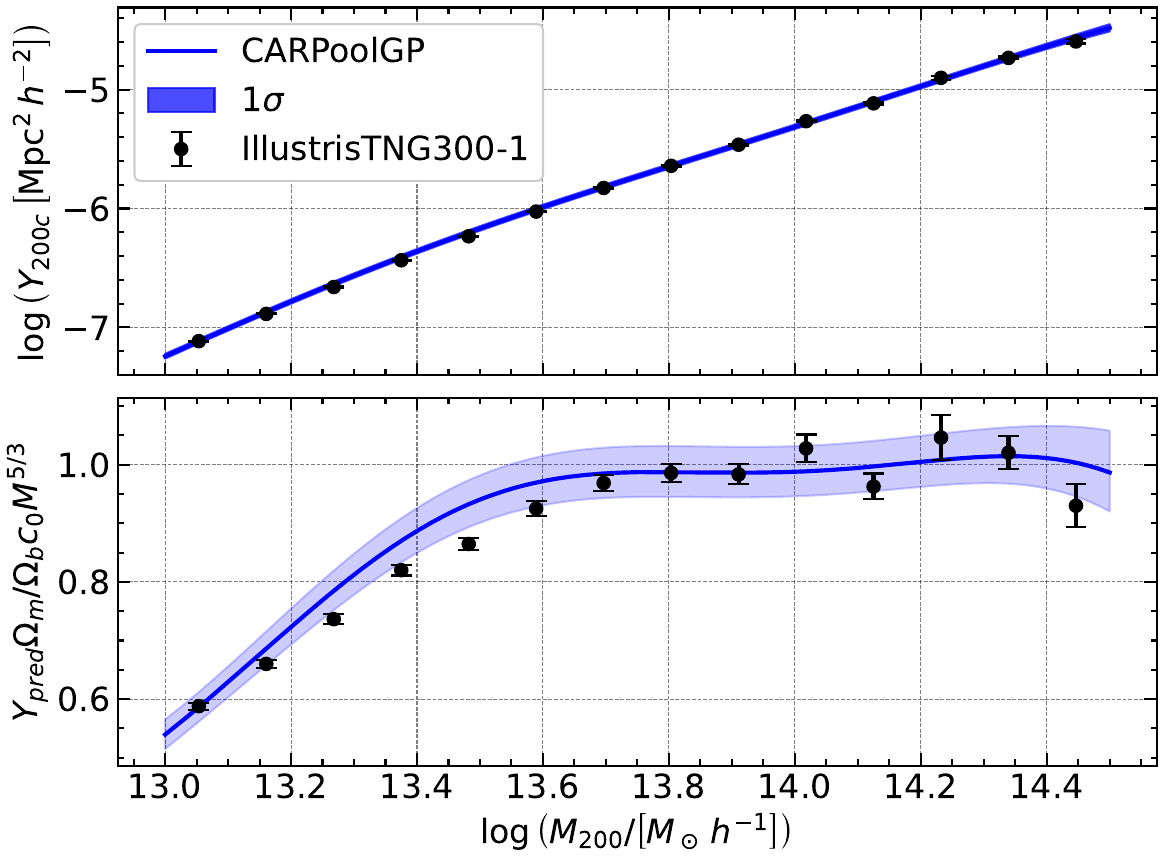}
    \caption{The fiducial parameters for IllustrisTNG300-1 are emulated using CARPoolGP (blue line) with their associated $1\sigma$ error bars and compared with the true values from IllustrisTNG (black points). The bottom panel shows the same emulation and comparison, but for the scaled $Y$ given by Eq.~\ref{eq:scaled_Y}. The discrepancy in emulated and IllustrisTNG predictions between $13.2\leq \log(M/[M_\odot\,h^{-1}])\leq 13.6$ contain sample fluctuations that are $\sim95\%$ correlated with their neighbors consistent with a low significance fluctuation.}
    \label{fig:Y_TNG}
\end{figure}

\subsection{Y-M emulations}\label{sec:emulations}

\begin{figure*}[t]
    \centering
    \includegraphics[width=\textwidth]{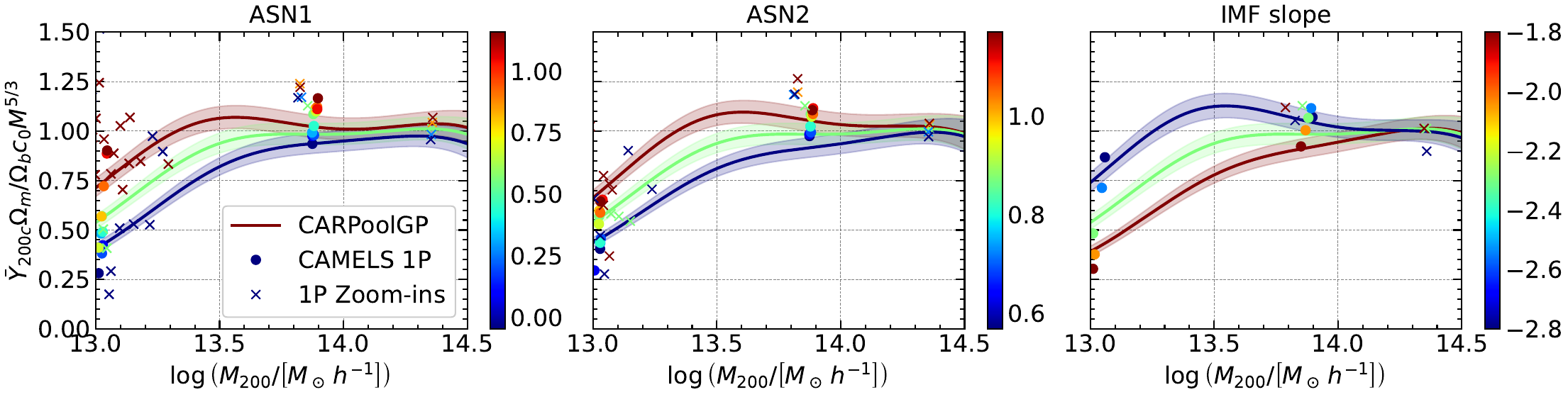}
    \caption{Individual parameter variations for the scaled $Y-M$ relation. We emulate each line with CARPoolGP trained on the full suite of zoom-in simulations, where all parameters are fixed to their fiducial values, while one parameter modulates between its bounds. Shaded regions represent the $1-\sigma$ predictive uncertainty. We explore in particular the stellar evolution and feedback parameters with the largest influence on the $Y-M$ relation. The CAMELS 1P set is incorporated into this plot using the largest halos in each box (dots), and its parameter values are colored accordingly. We include a set of two zoom-in halos, each with individual parameter variations (crosses), and a small subset of zoom-in halos run at the prior bounds of ASN1 and ASN2. See Fig.~\ref{fig:Y_emu_1P} for the full 28-dimensional emulation of the $Y-M$ relation.}
    \label{fig:Y_emu_1P_ASN_AGN}
\end{figure*}

Under the approximation that gas in the most massive halos is in hydrostatic equilibrium in their deep gravitational potential wells, a simple power-law relationship exists between the mass and integrated Compton Y parameter (Eq.~\ref{eq:y200}) $Y\propto M^{5/3}$ \citep{Kaiser-1986, Bryan-1998}. The amplitude of this power law, $c_0$ must be calibrated against the largest mass halos as, in detail, it depends upon the baryonic and total matter density profiles, as well as the gas temperature. We computed $c_0$ for IllustrisTNG by fitting the power law to halos with mass $\log(M_{200}/\left[M_\odot\,h^{-1}\right]) > 14$ and found $c_0=10^{-27.84} M_\odot^{-5/3} \Mpc^2\,h^{1/3}$.  It is common to rescale $Y$ as \citep{Wadekar-2023a},
\begin{equation}\label{eq:scaled_Y}
    Y_{200,c} = c_0M_{200}^{5/3}\left(\dfrac{\Omega_b}{\Omega_M }\right),
\end{equation}
where the right-hand side of the equation is especially useful for visualizing the $Y-M$ relation because it shows the deviation from the ideal self-similarity at lower masses as shown in \cite{Wadekar-2023a} and \cite{Pop-2022}. In Fig.~\ref{fig:Y_TNG}, the data points indicate the Y-M relation using $Y$ extracted from the IllustrisTNG300-1 simulation (top panel), and the scaled $Y-M$ relation using Eq.~\ref{eq:scaled_Y} (bottom panel). The error bars in both subpanels are Poissonian. 

Fig~\ref{fig:Y_TNG} provides a comparison of the IllustrisTNG300-1 results, which, thanks to the large simulation volume, can serve as a low-sample-variance reference for the fiducial IllustrisTNG model, with the CARPoolGP emulated results trained on the full suite of our zoom-in simulations. Specifically, we compare IllustrisTNG300-1 to the emulated $Y-M$ relation at parameter values matching the fiducial IllustrisTNG model. The IllustrisTNG300-1 halos are shown as black points with Poisson error bars while the emulated profile is shown as the blue line with its $1\sigma$ predictive uncertainty in the shaded band. It is worth noting that the resolution of our zoom-in simulations is similar to that of IllustrisTNG300-1, such that we ideally expect our emulator to reproduce the IllustrisTNG300-1 results closely.

In fact, there is an overall strong agreement between CARPoolGP and IllustrisTNG300-1 with a deviation from self-similarity in lower masses and a strong agreement in the highest masses. There appear to be four data points in tension with the IllustrisTNG result between $13.2 \leq \log(M_{200}/\left[M_\odot\,h^{-1}\right]) \leq 13.6$, however the emulated values of these points are $\sim 95\%$ correlated with one another implying that they can be represented as a single point in this mass range. The true tension then between the emulated and IllustrisTNG data in this region is small at the $\sim 2\sigma$ level. This result is promising evidence that CARPoolGP has learned enough to perform predictions across the entire mass range at the fiducial parameter values and match IllustrisTNG's results with a reduced variance estimate. We emphasize that the emulator was not trained on simulations at the fiducial IllustrisTNG parameter space coordinate, let alone the full mass range at this location.

In Fig.~\ref{fig:Y_emu_1P_ASN_AGN}, we modulate some parameters between the extrema of their prior bounds while keeping all other parameters fixed at their fiducial values. We focus on the three stellar processes-related parameters with the largest influence on the $Y-M$ relation while showing the $Y-M$ relation over the full 28-dimensional parameter space in Fig.~\ref{fig:Y_emu_1P} of Appendix~\ref{sec:appendix} (see \citetalias{Ni-2023} for a full description of the parameters). Fig.~\ref{fig:Y_emu_1P_ASN_AGN} shows that the $Y-M$ relation and its deviation from self-similarity are sensitive to stellar processes such as SN feedback (controlled by ASN1 and ASN2) and star formation (IMF slope). As the strength or velocity of SN winds, via ASN1 and ASN2, respectively, increases (decreases), we find an overall enhancement (suppression) of the $Y-M$ relation. Similarly, as the number of massive stars increases (decreases) with a shallower (steeper) IMF slope, we observe a suppression (enhancement) of the $Y-M$ relation. We discuss the physical causes of these effects below in Section~\ref{sec:physical_picture}.

In the flagship CAMELS $(25\,\Mpc\,h^{-1})^3$ volume boxes, a set of simulations that performed this exact method of individual parameter variation (the CAMELS 1P set) to explore this dependency at the level of cosmological boxes. A similar suite of individual parameter variation zoom-in simulations containing two halos and a small subset of zoom-in simulations run for ASN1 and ASN2 were generated for validation purposes. We add points extracted from the CAMELS 1P set as well as points generated from the 1P zoom-in simulations in Fig.~\ref{fig:Y_emu_1P_ASN_AGN} to gauge the emulator's performance, particularly at the lowest masses. We see that qualitatively, there is agreement between the emulated prediction of a parameter's evolution and the 1P set. When the emulated $Y-M$ relation produces the same results after modulating a parameter, this means that that parameter does not influence the $Y-M$ relation.

\begin{figure*}[t]
    \centering
    \includegraphics[width=0.98\textwidth]{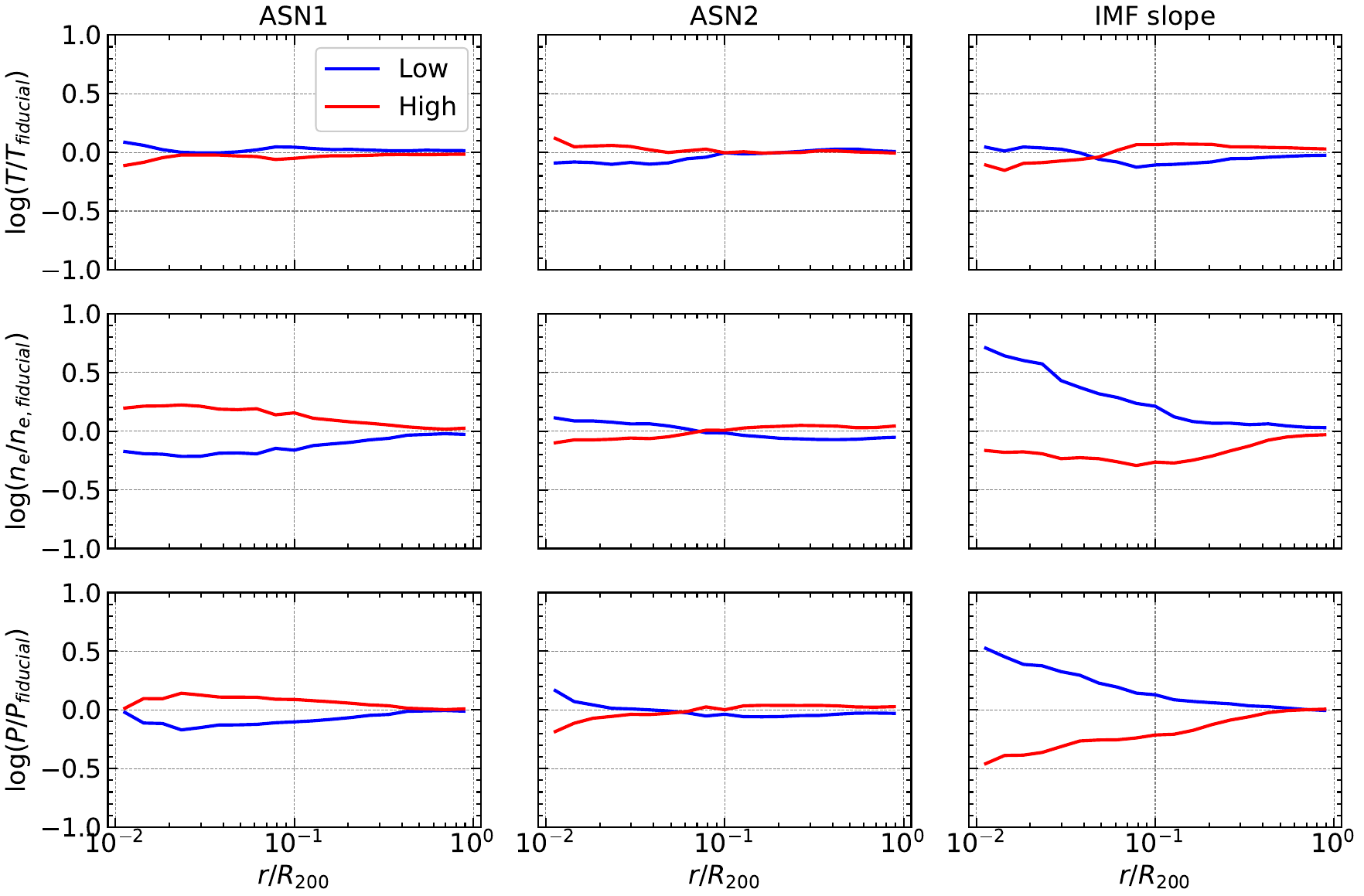}
    \caption{The emulated temperature (top row), electron density (middle row), and pressure (bottom row) profiles for a $\log(M_{200}/\left[M_\odot\,h^{-1}\right])=13.75$ halo as a function of radius in twenty log spaced bins between $0.01\leq r/R_{200}\leq 1$ at the extreme values for three astrophysical parameters. Each plot shows the profiles normalized by the fiducial to highlight their deviations. We find that the dominant source of pressure changes for each of the chosen parameters is due to changes in their density profiles. Changes in the temperature profiles of the gas are suppressed except for the IMF slope.}
    \label{fig:temp_density}
\end{figure*}

\subsection{A physical picture of parameter trends in the Y-M relation}\label{sec:physical_picture}

Two competing effects modify the pressure in the IGrM/ICM and, therefore, the $Y-M$ relation (Eq.~\ref{eq:y200}): SN winds originating from the ISM and AGN feedback from the central black hole. 

This section provides a qualitative physical picture of how the IllustrisTNG model parameters influence SN and AGN feedback in IGrM and ICM. This allows us to better understand the parameter trends in the $Y-M$ relation (see Fig.~\ref{fig:Y_emu_1P_ASN_AGN} and Fig.~\ref{fig:Y_emu_1P}). It is important to note that a rigorous investigation into the interplay between SN and AGN feedback in the IGrM/ICM is beyond the scope of this work, requiring more in-depth theoretical study and comparisons to observations. However, for the first time, we can observe how changes in the galaxy formation model parameters influence these processes and their subsequent effects on galaxy formation. We begin by introducing various emulations of physical quantities, which we then tie together into a physical picture of the IGrM and ICM. During this process, we discover puzzling effects that open the door to interesting avenues for future research.

\begin{figure*}[t]
    \centering
    \includegraphics[width=0.99\textwidth]{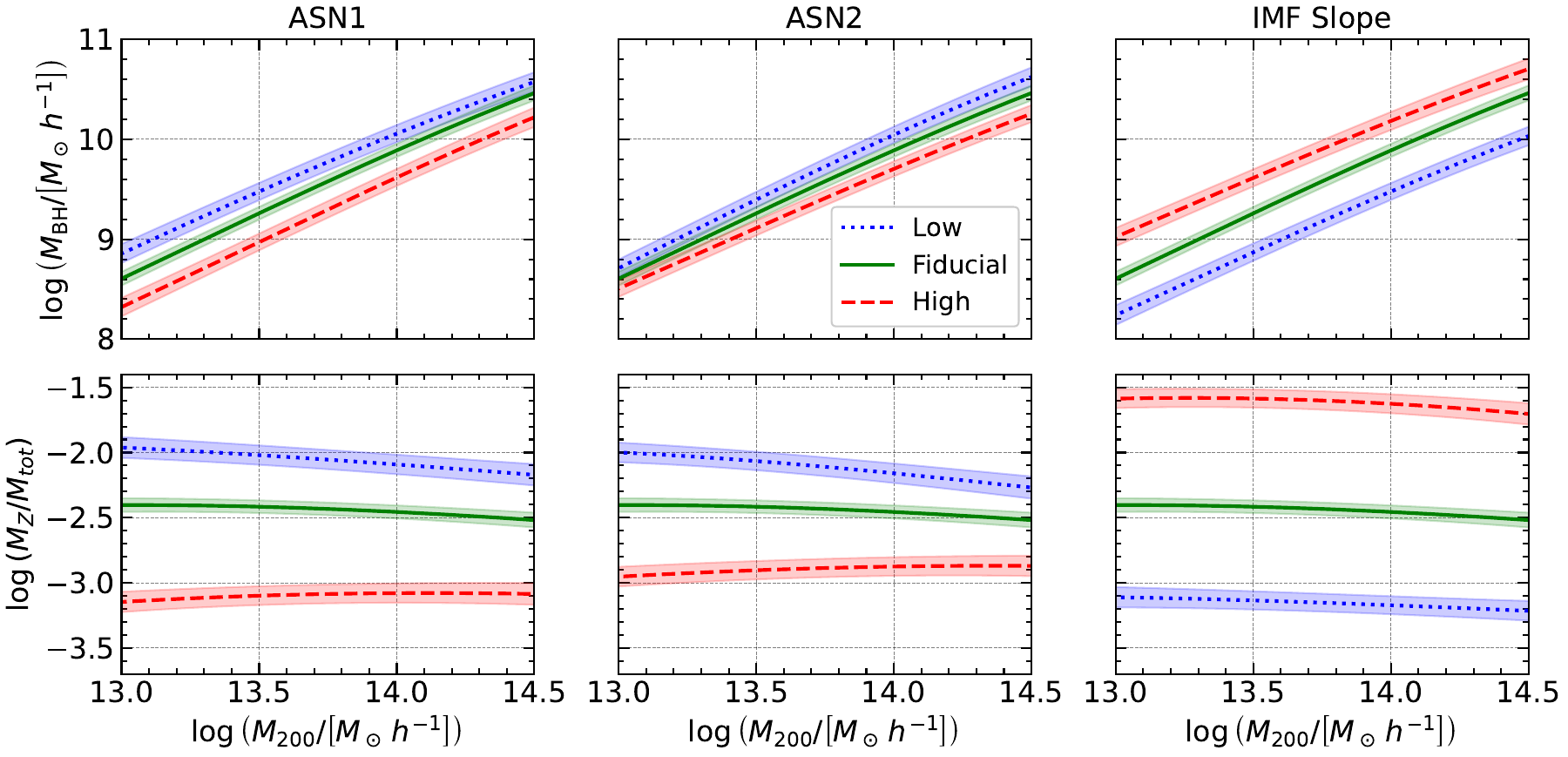}
    \caption{The emulated black hole mass (upper panels) and gas metallicity (lower panels) as a function of halo mass at the high (red), fiducial (green), and low (blue) bounds of ASN1 (left panels), ASN2 (center panels), and the IMF slope (right panels). We find a significant suppression (enhancement) of the central black hole mass with an increase (decrease) in the energy per unit mass of stars formed in SNe winds, ASN1. We speculate that this is due to an increased (decreased) mass loading factor (Eq.~\ref{eq:massloading}), which removes gas from the black hole's accretion supply. The increases in the IMF slope lead to a metallicity enrichment in the halo, which cools and collapses the gas and allows the central black hole to grow larger.}
    \label{fig:BHM_Metal}
\end{figure*}
\subsubsection{ASN1}

The model parameters showing dominant effects on the $Y-M$ relation are ASN1, ASN2, and the IMF slope. Here, ASN1 and ASN2 are related to the IllustrisTNG model's supernova feedback and represent normalization factors for the energy of galactic winds per star formation rate and velocity of galactic winds, respectively \citep{Pillepich-2018, Ni-2023}. The IMF slope is the slope of a modified \cite{Chabrier-2003} Initial Mass Function above $1M_\odot$.
 
In the IllustrisTNG model, SN feedback is parameterized by the mass loading factor \citep{Pillepich-2018}
\begin{equation}\label{eq:massloading}
    \eta \equiv \dfrac{\dot{M}_{W}}{ \dot{M}_{\star}} = \dfrac{2}{v_w^2}e_w(1-\tau_w),
\end{equation}
where $\dot{M}_{W}$ is the wind mass outflow rate, $\dot{M}_{\star}$ is the star-formation rate, $v_w$ is the redshift dependent SN wind velocity normalized by ASN2, $e_w$ is the metalicity-dependent energy per unit mass of formed stars in SN winds normalized by ASN1 and $\tau_w$ is the fraction of thermal energy given to SN winds.

AGN feedback in the IllustrisTNG model occurs in two main states: thermal and kinetic. In the thermal mode, the black hole is inefficient at heating or ejecting gas far away from the black hole, while in the kinetic mode, major ejection events occur that can efficiently heat the halo gas and even expel it out of the halo. The transition to the kinetic mode in the fiducial TNG model occurs when black holes are above the accretion rate threshold defined as \citep{Weinberger-2017}
\begin{equation}\label{eq:accretionthreshold}
    \chi = \min\left[\chi_0\left(\frac{M_{BH}}{10^8 M_\odot}\right)^\beta, 0.1\right],
\end{equation}
where $\chi_0$ is a threshold normalization (\texttt{QuasarThreshold}) and  $\beta$ is a power law scaling index (\texttt{QuasarThresholdPower}) \citep{Weinberger-2017, Ni-2023}. In the fiducial IllustrisTNG model, this transition occurs at a black hole mass of $10^8\,M_\odot$. Thus, any suppression or enhancement of the black hole's accretion can affect the thermodynamic state of the halo.

In Fig.~\ref{fig:temp_density}, we present the emulated temperature, electron density, and pressure profiles for a $\log(M_{200}/\left[M_\odot\,h^{-1}\right]) = 13.75$ halo for ASN1, ASN2, and the IMF slope. To generate this, we compute the mass-weighted temperature, the mean electron density, and electron pressure in twenty log-spaced bins between $0.01\leq r/R_{200}\leq 1$ from the suite of zoom-in simulations. We then train a CARPoolGP emulator for each bin. We use this set of CARPoolGP emulators to construct profiles at the highest and lowest bounds of ASN1, ASN2, and the IMF slope. Then, we normalize the profiles by emulations at the fiducial parameter space location in order to highlight the variations with respect to the fiducial model. 

\begin{figure*}[t]
    \centering
    \includegraphics[width=0.98\textwidth]{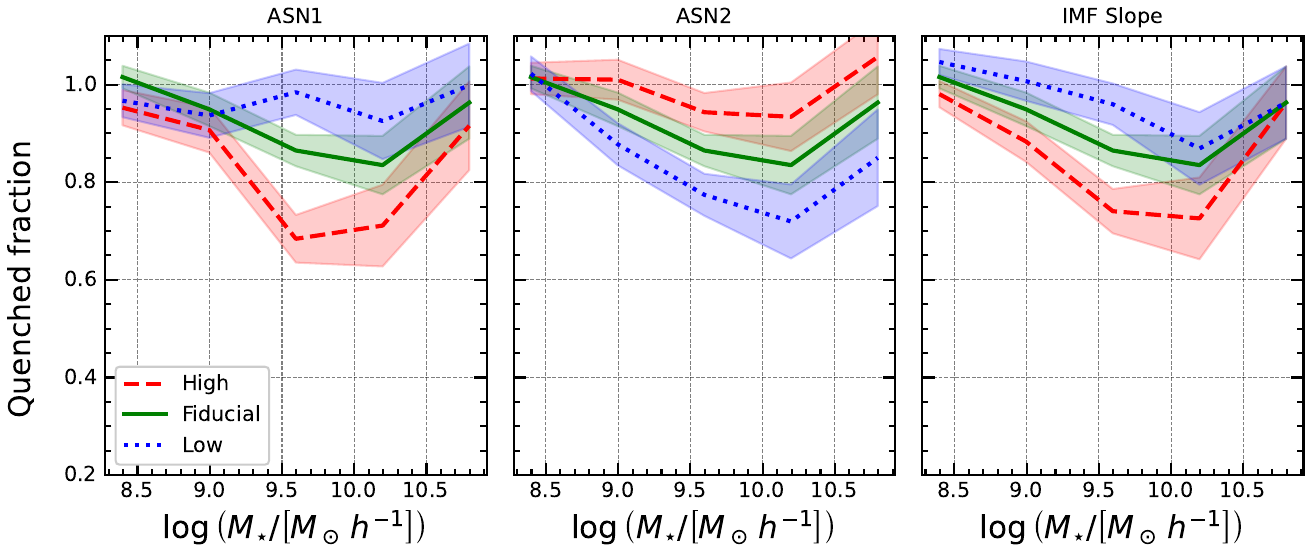}
    \caption{The quenched fraction of satellite galaxies in a $\log(M_{200}/\left[M_\odot\,h^{-1}\right]) = 13.75$ halo as a function of satellite stellar mass. The parameter variations are the same as in Fig.~\ref{fig:BHM_Metal}. Decreases in the black hole mass (Fig.~\ref{fig:BHM_Metal}) correspond to suppressed quenching of satellite galaxies for ASN1, while for ASN2 and the IMF slope, slowing the speed of stellar winds and enhancing the gas enrichment reduces the quenching of satellite galaxies.}
    \label{fig:quenchedSN}
\end{figure*}

For reasons that will soon become clear, we complement Fig.~\ref{fig:temp_density} with Fig.~\ref{fig:BHM_Metal} and Fig.~\ref{fig:quenchedSN} where we emulate the central black hole mass and gas metalicity as a function of halo mass, and the satellite galaxy quenched fraction for a $\log(M_{200}/\left[M_\odot\,h^{-1}\right])$ halo as a function of satellite stellar mass, respectively. We define the satellite galaxy quenched fraction in a stellar mass bin following \citet{Donnari-2021},
\begin{equation}
Q_i = \dfrac{1}{N_i} \sum_{j}^{N_i} q_{j}
\end{equation}
\begin{equation}
\begin{split}
    q_{j} = \begin{cases}
    1\ \ \ \ & \dot{M}_{\star, j}/M_{\star, j} \leq 10^{-11}\,{\rm yr}^{-1}\\
    0\ \ \ \ &\dot{M}_{\star, j}/M_{\star, j} > 10^{-11}\,{\rm yr}^{-1},
\end{cases}
\end{split}
\end{equation}
where $N_i$ is the number of satellite galaxies in the $i$-th stellar mass bin, $\dot{M}_{\star, j}$ is the sum of star formation rates within twice the stellar half radius $R_{1/2, j}$ of satellite $j$, $M_{\star,j}$ is the total stellar mass within $R_{1/2, j}$, and the sum is performed over all satellites within the appropriate stellar mass bin. We consider five logarithmic spaced bins between $8\leq M_\star/M_\odot \leq 11$, and we only consider satellites inside $R_{200}$.

As ASN1 increases, the energy of the winds increases along with the mass loading factor. We expect a decrease in the star formation rate as stellar winds can transfer more energy and mass out of the Interstellar Medium (ISM) and into the IGrM and ICM. This, in turn, reduces the metallicity of the IGrM and ICM (Fig.~\ref{fig:BHM_Metal}, lower left panel) and restricts the growth of the central black hole (Fig.~\ref{fig:BHM_Metal} upper left panel). 

The delayed transition into the kinetic mode can reduce the influence of the AGN feedback compared to the fiducial. The combination of an increased mass loading factor that removes more gas from the ISM, decreased metallicity, which limits the cooling of gas and star formation, and delayed BH growth, which restricts the powerful expulsion of gas from the halo, leads to a net increase in the density throughout the halo (Fig.~\ref{fig:temp_density} left center panel). Further, we expect the reduced black hole growth to limit the powerful AGN quenching of surrounding satellite galaxies. We see this to be true in the left panel of Fig.~\ref{fig:quenchedSN}. The overall temperature (Fig.~\ref{fig:temp_density} left upper panel) of the halo is not greatly affected by these changes as expected \citep{Loken-2002}. The effects on the pressure profile (Fig.~\ref{fig:temp_density} left lower panel) appear to be dominated by the density changes. A similar, mirrored story can be told for the decrease in ASN1. The decrease in mass loading allows the central black hole to accrete faster and enter the kinetic feedback mode at earlier times. This efficient feedback mode displaces gas throughout the halo and decreases density at all radii while efficiently quenching satellite galaxies in the process. 

In this qualitative picture, we speculate that the dominant mechanism affecting the pressure is the AGN feedback, even though the suppression/enhancement of the central black hole growth via SN winds spawns this effect. Strangely, we find that the AGN parameters, which can also affect the growth of the central black hole, do not give rise to significant changes to the $Y-M$ relation. This is somewhat in tension with the above picture and implies either an alternative mechanism controlling the density of the IGrM and ICM or a complex interaction between the SNe and AGN feedbacks, which we have overlooked. We leave this as an exciting avenue for future research. 

\subsubsection{ASN2}
For ASN2, as the velocity of stellar winds is enhanced, wind particles deposit their energy farther out in the halo and, similar to ASN1, decrease the star formation rate. This leads to a similar effect to ASN1, reducing the metallicity of the gas (Fig.~\ref{fig:BHM_Metal}, center lower panel) and starving the black hole (center upper panel). However, it is clear from Fig.~\ref{fig:BHM_Metal} that this effect is smaller when compared to ASN1. In fact, the density profiles of Fig.~\ref{fig:temp_density} (center center panel) show what we would expect from an increase in SN wind velocity in the absence of kinetic AGN feedback. Namely, we find that increasing ASN2 depletes the ISM of gas moving it to the outer regions of the halo. This causes an increase in the quenching of satellite galaxies (Fig.~\ref{fig:quenchedSN}, center panel) as their star-forming material is more removed and placed at farther stretches in their CGM, less able to recycle back and contribute to further star-formation.

SN-driven winds are generally not powerful enough to expel wind above the virial velocity of these halos, as can kinetic AGN feedback. In this way, we find that for a high-velocity wind scenario, there are denser regions in the outer IGrM and ICM, whereas, in low-velocity scenarios, the density is enhanced in the inner $10\%$ of the halo. We find a slight increase in the average gas temperature in the inner $10\%$ of the halo but a mostly constant temperature profile throughout the rest of the halo. Therefore, the dominant effect on the pressure is again due to changes in density in this case. We can conclude from this that the observed effects in the $Y-M$ relation of Fig.~\ref{fig:Y_emu_1P_ASN_AGN}, are sourced by the changes of the SN feedback such that increases in the SN winds enhance the density and pressure of the IGrM which leads to an enhanced $Y-M$ relation.

\subsubsection{IMF Slope}
The IMF slope controls the distribution of the masses of individual stars. As the IMF slope increases (becomes less negative), more massive stars are formed, which implies more supernovae and, hence, more chemical enrichment and SN feedback. On the one hand, we expect an increase in the IMF slope and, hence, in the number of supernovae to give rise to effects similar to those seen in ASN1: a reduction in the growth of the black hole and depletion of the gas. Instead, we find almost the exact opposite. The reason for this is the IMF slope's major effect on the gas's metallicity (Fig.~\ref{fig:BHM_Metal}, lower right panel). The increase in chemical enrichment in the TNG model leads to a reduction in supernova efficiency \citep{Pillepich-2018}, but also to an improvement in the cooling function \citep{Vogelsberger-2013}. The heavy ions lead to an enhanced cooling flow in the halo, allowing the central black hole to grow significantly faster (Fig.~\ref{fig:BHM_Metal}, upper right panel). The faster transition into the kinetic AGN mode ejects more gas than in the fiducial and low IMF slope cases. 

To further support this picture, we show that the quenching of satellite galaxies (Fig.~\ref{fig:quenchedSN} right panel) is reduced as the IMF slope increases. The increased cooling from the enrichment of the IGrM and ICM leads to an enhancement of star formation in satellite galaxies. While the satellites may still experience environmental quenching from the central black hole in the kinetic feedback mode, their own black holes have likely not grown large enough to start internal quenching processes. Thus, we see that star formation due to the increased reservoir of cold gas in the satellites dominates over the environmental quenching mechanisms.

\section{Discussion}\label{sec:discussion}
\defcitealias{Wadekar-2023b}{W23}

In this section, we first compute Fisher matrix constraints on the astrophysical parameters for the $Y-M$ relation. We show that by using high-mass halos, the predicted constraints from next-generation SZ experiments are significantly tightened compared to \cite{Wadekar-2023b} (henceforth \citetalias{Wadekar-2023b}), but that when marginalizing over the entire IllustrisTNG parameter space, these constraints significantly diminish. We then discuss the caveats in our CARPoolGP testing methods.

\subsection{Astrophysical parameter constraints}\label{sec:Fisher}

\begin{table}[]
    \centering
    \begin{tabular}{|c|c|}
    \hline
        $\log_{10}(M)\,[M_{\odot}]$  &  $\log_{10}(Y^{+ 1\sigma}_{-1\sigma})\,[\Mpc^2\,h^{-2}]$\\
        \hline
        $13 - 13.5$ & $-6.584^{+0.02}_{-0.015}$\\
        $13.5 - 14$ & $-5.748^{+0.007}_{-0.006}$\\
        $14 - 14.5$ & $-4.902^{+0.004}_{-0.005}$ \\
        \hline
    \end{tabular}
    \caption{Forecasted constraints on the $Y-M$ relation for CMB-S4 and DESI like surveys. See \cite{Pandey-2023} for more details.} 
    \label{tab:YM_forecast}
\end{table}
In Fig.~\ref{fig:Y_emu_1P_ASN_AGN}, we showed that astrophysical parameters modify the magnitude and deviation from self-similarity in the $Y-M$ relation. In this section, we consider the possibility of constraining the strength of astrophysical parameters with a CMB-S4 and a DESI-like survey. In \citetalias{Wadekar-2023b}, the strength of four astrophysical parameters was constrained using the CAMELS $(25\,\Mpc\,h^{-1})^3$ boxes containing halos with $\log(M_{200}/\left[M_\odot\,h^{-1}\right])<13$. Here, we replicate the analysis but use halos with $13\leq\log(M_{200}/\left[M_\odot\,h^{-1}\right])\leq14.5$ based on our zoom-in simulations and CARPoolGP, and observational constraints in larger mass bins. Just as in \citetalias{Wadekar-2023b}, we use forecasts of the $Y-M$ relation following \citet{Pandey-2023}. We refer the reader to \citet{Pandey-2023} for details on the generation of the $Y-M$ relation, and show in Table~\ref{tab:YM_forecast} the forecasted constraints for three log mass bins between $13\leq \log(M_{200}/\left[M_\odot\,h^{-1}\right])\leq 14.5$ (\citet{Pandey-2023} private communication). 

\citetalias{Wadekar-2023b} solely had access to four astrophysical parameters (WindEnergyin1e51erg - ASN1, RadioFeedbackFactor - AGN1, VariableWindVelFactor - ASN2, RadioFeedbackReorientationFactor - AGN2). To provide a similar comparison, we limit the astrophysical parameter space to include these same four parameters. We perform a Fisher forecast assuming the parameter distributions are Gaussian, and compute \citep{Tegmark-1997}:
\begin{equation}\label{eq:Fisher}
    F_{ab} = \sum_{ij}^{max} \dfrac{\partial \log \bar{Y}_i}{\partial \log \theta_a} C_{ij}^{-1}\dfrac{\partial \log \bar{Y}_j}{\partial \log \theta_b},
\end{equation}
where $\bar{Y}_{i}$ is the mean $Y$ for a given mass bin, $\theta$, is the astrophysical parameter, and $C^{-1}$ is the inverse covariance matrix. 

CARPoolGP is equipped to emulate $Y$ at any mass, so we treat $\log(\bar{Y})$ as an emulation at the center of each mass bin ($\log({M_{200}/\left[M_\odot\,h^{-1}\right]}) = [13.25, 13.75, 14.25]$), given the fiducial set of IllustrisTNG parameters. We have written CARPoolGP in the JAX programming language, which offers auto-differentiation of input variables. This allows us to compute directly the derivatives with respect to astrophysical parameters. Finally, for the covariance matrix, we use the values in Table~\ref{tab:YM_forecast}, and assume that the mass bins are uncorrelated so that the matrix is diagonal. We generate the matrix elements using the mean $Y$ values and errorbars such that
\begin{equation}
    C_{ii} = \left(\dfrac{\Delta Y_i}{\bar{Y}_i}\log(e)\right)^2,
\end{equation}
where $\Delta Y_i$ is one-half the width of an error bar on $Y$ shown in Table~\ref{tab:YM_forecast}. Just as in \citetalias{Wadekar-2023b}, we center the covariance measurements on the emulated values of $Y$ but find minimal effects when using the mean values provided in Table~\ref{tab:YM_forecast}. 

\begin{figure}[t]
    \centering
    \includegraphics[width=0.48\textwidth]{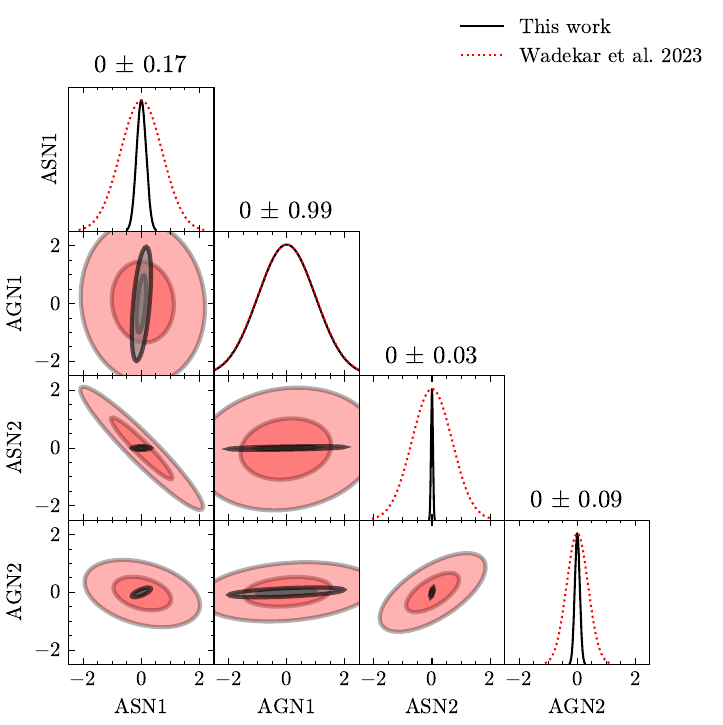}
    \caption{A Fisher forecast, using our high-mass zoom-in simulation suite, on the four astrophysical parameters used in the flagship CAMELS $(25\,{\rm Mpc}\,h^{-1})^3$ boxes. We generate derivatives and mean $\log Y$ values using CARPoolGP and its autodifferentiability, along with forecasted constraints on the Y-M relation from \cite{Pandey-2023}. The Fisher forecast using low mass halos in \citetalias{Wadekar-2023b} are shown as red contours and lines. We find strong constraints on each astrophysical parameter, except for AGN1.}
    \label{fig:Fisher_Y}
\end{figure}

To ensure that the Fisher matrix is invertible when using three mass bins but four parameters, we add a weak Gaussian prior, $\sigma_{\log \theta} = 1$, to the diagonal elements. We then invert this to find $F^{-1}_{ab}$ and construct Gaussian constraints. We show the results in Fig.~\ref{fig:Fisher_Y}. Our results are a significant improvement over those shown in Fig. 6 of \citetalias{Wadekar-2023b}. We find that, apart from AGN1, all astrophysical parameters are strongly constrained, at up to an order of magnitude in constraining power. One reason for this is that the error on the $Y-M$ relation is much smaller at the highest masses, allowing for more sensitivity to deviations in the $Y-M$ relation. Here, we leverage the accuracy of the $Y-M$ observations from Table~\ref{tab:YM_forecast} to obtain such tight constraints. 

\begin{figure}[t]
    \centering
    \includegraphics[width=0.48\textwidth]{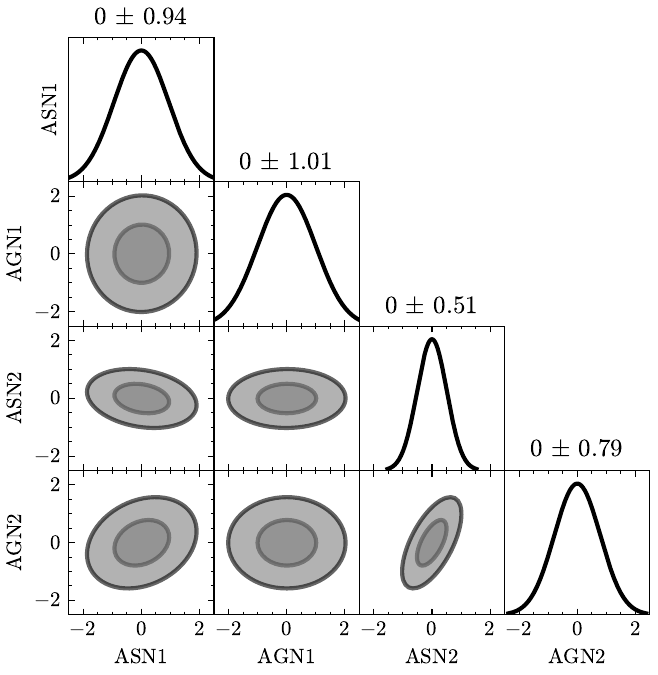}
    \caption{Exactly the same as Fig.~\ref{fig:Fisher_Y}, but after marginalizing over the full 28-dimensional IllustrisTNG parameter space. }
    \label{fig:Fisher_YM}
\end{figure}

With access to the 28-dimensional IllustrisTNG model parameter space, we can extend the analysis of \citetalias{Wadekar-2023b}. In Fig.~\ref{fig:Fisher_YM}, we show constraints on the same astrophysical parameters after marginalizing over the entire parameter space. We follow the same process as the preceding paragraphs and with the same covariance matrix from Table~\ref{tab:YM_forecast}. We find that the constraints are greatly reduced. This highlights the importance of including the full set of parameters when constraining model parameters. Furthermore, it shows that with only three mass bins in the $Y-M$ relation, we cannot provide tight constraints and instead would require more observations or complementary probes.

\subsection{Testing CARPoolGP}\label{Tests}
CARPoolGP was designed to emulate astrophysical quantities across a high-dimensional astrophysical and cosmological parameter space, extract parameter dependencies for given quantities, and help to determine the best locations to draw future samples. Testing its performance in this context is difficult, as there is no existing suite of simulations with a similar parameter space exploration. In this work, we have used comparisons with existing simulations such as IllustrisTNG300-1 \citep{Pillepich-2018}, a limited sample of CAMELS boxes \citep{Villaescusa-Navarro-2021} (Section~\ref{sec:emulations}), or some internal measure of variance reduction (Section~\ref{sec:sampling_stages}). Although each test is instructive, it is essential to discuss its caveats. 

The most straightforward test we perform is an internal consistency test (Fig.~\ref{fig:AL_hist}), where we use Eq.~\ref{eq:CPprediction} to predict the mean and variance of some quantity at parameter space locations that have not been used in the simulation suite. We use this test to better understand the parameter space locations that provide a global reduction in variance and to explore the efficacy of each active learning step. We found that the variance in the predictions decreased effectively after each of the four stages, with the latter stages performing better than expected. However, testing this way does not provide insight into a potentially biased estimator. It could be that each step internally reduces the global variance, but ultimately, the emulator maintains some level of biased predictions. It is unclear if a bias could even occur due to this sampling strategy, but this cannot be ruled out without further testing. In some ways, this can be mitigated by using the base samples of stage 1, which are drawn from a Sobol sequence. This would reduce any potential bias but result in an increased predictive variance.  

In Fig.~\ref{fig:Y_TNG}, we compare the emulated results with IllustrisTNG300. We found strong agreement between the emulated results and IllustrisTNG throughout the full mass range, with a small tension between $13.2\leq \log(M_{200}/\left[M_\odot\,h^{-1}\right])\leq 13.6$. We noted that sample variance fluctuations in this mass range were highly correlated, and therefore, the discrepancy between the emulated predictions and the IllustrisTNG results is consistent with a low-significance fluctuation. While we expect the CARPoolGP emulator to provide a low-variance description across the entire space, we note that this emulation cannot be sensitive to large quantity fluctuations that may exist between very nearby parameter space locations. This is the case for any emulator or interpolation with limited training data. Furthermore, testing CARPoolGP at a single location in the parameter space does not provide any information on its ability to capture specific parameter dependencies or degeneracies. These degeneracies could be necessary for scenarios where one is interested in performing inference on some survey data set to obtain the best-fit astrophysical and cosmological parameters, or if one wants to marginalize over baryonic effects effectively, knowledge of the individual parameter dependencies will prove critical. While testing CARPoolGP's ability at the fiducial set of parameters does not explore these dependencies, it does provide an intuitive demonstration of CARPoolGP's capabilities. 

We show CARPoolGP's ability to capture parameter dependencies by comparing emulations to previous CAMELS simulations and find a strong agreement between the emulated parameter dependencies and the simulated parameter dependencies (Fig.~\ref{fig:Y_emu_1P_ASN_AGN} and Fig.~\ref{fig:Y_emu_1P}). However, we must consider two caveats when we explore individual parameter variations with the CAMELS 1P set. First, the flagship CAMELS box is small at $(25\,\Mpc\,h^{-1})^3$, while the suite of zoom-in simulations is generated in a large, $(200\,\Mpc\,h^{-1})^3$ box. The zoom-in simulations are, therefore, generated with large-scale modes that the small boxes cannot access. These largest scales can affect the formation of galaxy groups and clusters and provide different results compared to small boxes when extracting summary statistics from them. It is then difficult to discern whether any tension between the small CAMELS boxes is due to the scale differences or sample variance. Second, there are not enough high-mass halos within these boxes for a robust comparison. Although some halos of mass $10^{13}\,M_\odot$ exist, this can only provide constraints on the lowest mass predictions from CARPoolGP. We also include in Fig.~\ref{fig:Y_emu_1P_ASN_AGN} two zoom-in clusters at two mass scales that were simulated with single-parameter variations at a time, but here it is clear that a single halo contains a great deal of sample variance, again, making any comparison difficult. Finally, while the CAMELS 1P set, which contains individual parameter variations, can assist us in exploring parameter dependencies, we run into a similar issue as described above when testing on IllustrisTNG300-1.

\section{Conclusions} \label{sec:conclusion}
This work introduces the CARPoolGP sampling and regression method as well as a suite of group-to-cluster mass zoom-in simulations and utilizes these tools to study the SZ effect in terms of the Compton relation $Y-M$ and additional scaling relations within the framework of the IllustrisTNG galaxy formation model. We highlight the key conclusions in the following.
\begin{itemize}
    \item We developed the general formalism for a novel reduced variance sampling and regression method called CARPoolGP. In CARPoolGP, correlations between quantities across a parameter space can be leveraged to provide high accuracy and low variance emulations. The CARPoolGP structure places one set of samples throughout the parameter space and a correlated set of samples at or near predefined locations in the parameter space to calibrate the predictions and variance effectively. 
    \item We showed that in a simple one-dimensional problem, CARPoolGP outperforms a random sampling approach, providing more accurate and reduced variance emulations. With CARPoolGP, we achieved $\sim 45\%$ better variance reduction than the case of random sampling in a 1-dimensional toy example.
    \item We built an active learning algorithm that predicts the best parameter space locations to draw future samples and which reduces variance on some quantity of interest across the entire parameter space. We found that this approach reduced the variance by $\sim65\%$ when compared to the random sampling approach and $\sim36\%$ amount compared to the standard CARPoolGP approach in a 1-dimensional toy example.
    \item We used CARPoolGP and the active learning approach to generate a suite of 768 hydrodynamical and analogous dark matter-only simulations spanning a 28-dimensional astrophysical and cosmological parameter space in the IllustrisTNG model, as well as a mass range $13\leq \log(M_{200}/\left[M_\odot\,h^{-1}\right])\leq 14.5$. We used active learning to enhance the variance reduction on the Compton $Y$ parameter.
    \item We trained a CARPoolGP emulator on the $Y-M$ relation. Our emulator predicted the fiducial IllustrisTNG $Y-M$ relation to high accuracy with reduced variance and matched expectations when emulating individual parameter variations and comparing them to existing simulations. 
    \item We explored the emulated thermodynamic profiles of halos along individual parameter variations, finding that the normalization of supernova wind energy per star formation rate (ASN1), normalization of supernova wind speed (ASN2), and IMF slope above $1\, M_\odot$ provide the largest changes to the $Y-M$ relation, while remarkably, the AGN feedback parameters have little impact on this relation.
    \item We developed a qualitative physical picture of gas in the IGrM and ICM using emulations of various thermodynamical profiles, metallicity, and black hole mass - halo mass relations, and satellite quenched fraction relations to explain the observed trends in the $Y-M$ relation due to ASN1, ASN2, and IMF slope.
    \item  We used the auto-differentiable capabilities of CARPoolGP and constraints on future SZ experiments to perform a Fisher forecast on four IllustrisTNG parameters. We found tighter constraints, by an order of magnitude, on three of the four astrophysical parameters compared to previous CAMELS studies, yet further showed that marginalizing over the full parameter space significantly weakens these constraints. 
\end{itemize}

We highlight that in this work, we applied CARPoolGP to simulate group and cluster mass halos, but this is just one use case. Future astrophysical studies considering the exploration of a model's parameter space, and with prior knowledge of correlations across this space, can benefit greatly from using CARPoolGP. While more research is required to investigate alternative architectures, such as multiple surrogates per base setups, the current ``out of the box" implementation is already beneficial and ready for use. 

Further, with the novel suite of high-mass halos spanning the IllustrisTNG model parameter space, a wide range of science applications are possible. We foresee that this suite will aid in developing galaxy formation models and allow for more robust cosmological analyses.

%% compile with \acknowledgment, you will get an error print to the screen
%% Also note that the acknowledgment environment does not support long amounts of text. If you have a lot of people and institutions to acknowledge, do not use this command. Instead, create a new \section{Acknowledgments}.
\section{Acknowledgments}
% \begin{acknowledgments}
This work was supported by the Simons Collaboration on “Learning the Universe.” The Flatiron Institute is supported by the Simons Foundation.
MEL is supported by NSF grants DGE-2036197 and AST-2108678. DN acknowledges support from the NSF grant AST 2206055. G.L.B. acknowledges support from the NSF (AST-2108470, ACCESS), a NASA TCAN award, and the Simons Foundation. DAA acknowledges support by NSF grants AST-2009687 and AST-2108944, CXO grant TM2-23006X, JWST grant GO-01712.009-A, Simons Foundation Award CCA-1018464, and Cottrell Scholar Award CS-CSA-2023-028 by the Research Corporation for Science Advancement.
% \end{acknowledgments}


\vspace{5mm}
% \facilities{}


% \software{}



\appendix

\section{Y-M relation across full parameter space}\label{sec:appendix}
\begin{figure*}[t]
    \centering
    \includegraphics[width=\textwidth]{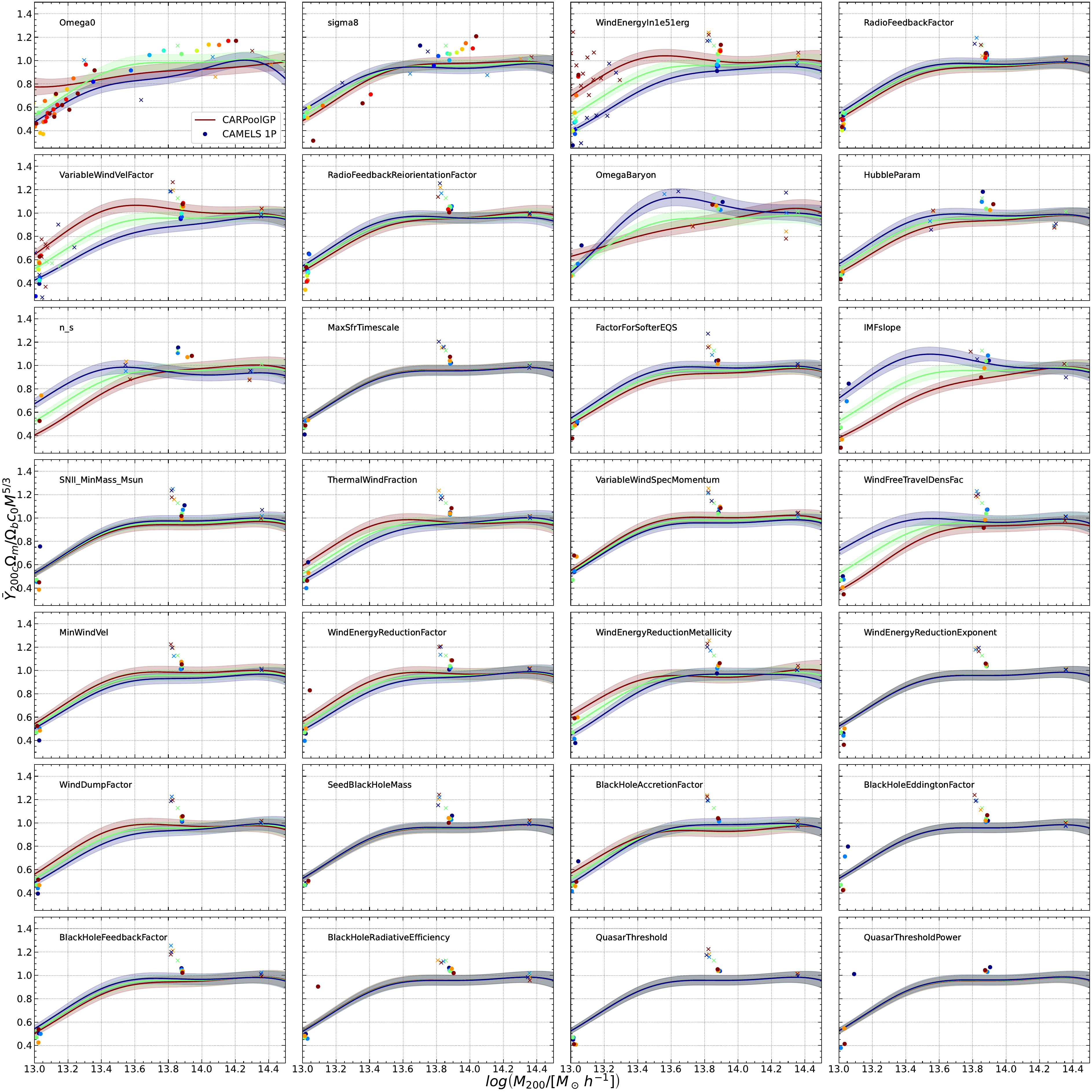}
    \caption{We present individual parameter variations for the scaled $Y-M$ relation. CARPoolGP emulates each line where all parameters are fixed to their fiducial values, while one parameter modulates between its bounds----Add high low. Shaded regions represent the $1-\sigma$ errors on predictions. The CAMELS 1P set is incorporated into this plot using the largest halos in each box, and their parameter values are colored accordingly.}
    \label{fig:Y_emu_1P}
\end{figure*}

We present in Fig.~\ref{fig:Y_emu_1P} emulations of the $Y-M$ relation for individual parameter variations across the entire parameter space of the IllustrisTNG model. We add this as a complement to Fig.~\ref{fig:Y_emu_1P_ASN_AGN}, where we selected three SN parameters providing the largest impact on the $Y-M$ relation and a few AGN parameters. We show this to allow readers to draw their own interpretations from the full set of parameters and further show the efficacy of CARPoolGP at emulating across the high dimensional parameter space. 

We find a weak effect on the thermodynamic properties of halos from the AGN feedback parameters in the IllustrisTNG model. This has been shown in multiple CAMELS-based works and matches our expectations \citep{Moser-2022, Wadekar-2023a}. Similarly, \citet{Singh-2022} and \citet{Tillman-2023} found that when exploring the six-dimensional parameter space of the original $(25\,\Mpc\,h^{-1})^3$ CAMELS boxes, the most significant effects on the multiphase CGM and intergalactic medium (IGM) occurred with modulations to the supernova parameter ASN2. Note that both \citet{Singh-2022} and \citet{Tillman-2023} speculated that the effects of ASN2 on the CGM and IGM were due to the complex interactions between SN winds and AGN feedback, such that changes in the SN parameters affected the growth of black holes and the power of the AGN feedback. With the high halo mass range of the mass function and the expanded parameter space, we find new parameter dependencies, such as a significant impact made by the IMF slope.

\bibliography{biblio.bib}{}
\bibliographystyle{aasjournal}



\end{document}